\documentclass[preprint]{aastex}
\usepackage{graphicx}        
\usepackage{natbib}         
\usepackage{amssymb}        
\usepackage{amsmath}
\usepackage{color}           
\usepackage{url}             
\usepackage{epstopdf}
\usepackage{soul}

\renewcommand{\vec}[1]{ {\mathbf #1} }

\begin{document}

\pagestyle{empty} 


\title{Fingering convection induced by atomic diffusion in stars: 3D numerical computations and applications to stellar models}
\author{Varvara Zemskova \\ Department of Marine Sciences, University of North Carolina at Chapel Hill, \\ 3202 Venable Hall, CB 3300, Chapel Hill, NC 27599-3300  \and  Pascale Garaud, \\ Department of Applied Mathematics and Statistics, Baskin School of Engineering, \\University of California at Santa Cruz, 1156 High Street, Santa Cruz CA 95064. \and Morgan Deal$^{1,2}$ \\ $^1$ Institut de Recherche en Astrophysique et Plan\'etologie, 14 avenue Edouard Belin, \\ Universit\'e de Toulouse, F-31400-Toulouse, France \\ $^2$ Laboratoire Univers et Particules de Montpellier, Universit\'e de Montpellier II, \\ Place Eug\`ene Bataillon, 34095 Montpellier C\'edex 05 FRANCE  \and Sylvie Vauclair \\ Institut de Recherche en Astrophysique et Plan\'etologie, 14 avenue Edouard Belin, \\ Universit\'e de Toulouse, F-31400-Toulouse, France}
\maketitle


\vspace{1cm}
\centerline{\bf Abstract} 
Iron-rich layers are known to form in the stellar subsurface through a combination of gravitational settling and radiative levitation. Their presence, nature and detailed structure can affect the excitation process of various stellar pulsation modes, and must therefore be modeled carefully in order to better interpret {\it Kepler} asteroseismic data. In this paper, we study the interplay between atomic diffusion and fingering convection in A-type stars, and its role in the establishment and evolution of iron accumulation layers. 
To do so, we use a combination of three-dimensional idealized numerical simulations of fingering convection, and one-dimensional realistic stellar models. Using the three-dimensional simulations, we 
first validate the mixing prescription for fingering convection recently proposed by Brown et al. (2013), and identify what system parameters (total mass of iron, iron diffusivity, thermal diffusivity, etc.) play a role in the overall evolution of the layer. We then implement the Brown et al. (2013) prescription in the Toulouse-Geneva Evolution code to study the evolution of the iron abundance profile beneath the stellar surface.  We find, as first discussed by Th\'eado et al. (2009), that when the concurrent settling of helium is ignored, this accumulation rapidly causes an inversion in the mean molecular weight profile, which then drives fingering convection. The latter mixes iron with the surrounding material very efficiently, and the resulting iron layer is very weak. However, taking helium settling into account stabilizes the iron profile against fingering convection, and a large iron overabundance can accumulate. The opacity also increases significantly as a result, and in some cases ultimately triggers dynamical convection.


\section{Introduction}
\subsection{The stellar context}

Atomic diffusion, a microscopic process which leads to the gradual spatial segregation of various chemical species, was recognised by the pioneers of stellar physics \citep[]{chapman17, chapman22, eddington26book} as one of the fundamental processes working in stellar interiors, and is a  straightforward consequence of the fact that stars are self-gravitating, pressure-supported gaseous spheres.

Indeed, as a first approximation, stars are in hydrostatic equilibrium, which means that the average weight of a fluid particle is balanced by the local pressure gradient. Meanwhile energy is transferred from the stellar core to the outer layers by radiation, except in convective zones where advection by fluid  motions is more efficient and therefore preferred. Radiative regions are assumed to reach a state of radiative  equilibrium, in which the energy transfer is governed by the local average opacity, which represents the absorption coefficient of radiation by the matter. As a consequence of this process, the local medium is also subject to an outward radiative pressure due the fact that the momentum transfer from the photons to the ions is slightly anisotropic. These physical considerations lead to the derivation of the fundamental stellar equations, which are then solved to compute so-called ``standard stellar models". 

However, as recognised since the 1920s, this standard treatment does not adequately take into account the fact that stars are made of {\it multicomponent} gases. In reality, these various components can behave quite differently in the presence of structural gradients. This manifests itself in several ways. The two most important effects are related to the different atomic weights of the various species, and to their different behavior in the process of radiative transfer. On the one hand, elements which have an atomic weight larger than the average are not entirely supported by the pressure gradient, and are therefore not in hydrostatic equilibrium when taken individually. Should this process act alone, they would gradually fall down towards the stellar center, while hydrogen (the lightest element) would slowly migrate upwards to take their place. This process is generally referred to as gravitational settling. On the other hand, it is clear that the ``average opacities" used in the computations of radiative transfer are also an approximation: in reality, each ion absorbs photons according to its own atomic structure and is consequently pushed up in a selective way. This process, referred to as ``radiative levitation" is thus inherently element-specific.


Depending on the relative amplitude of the  gravitational and radiative effects, with some added processes related to the thermal and concentration gradients, different atomic species move up or down during a collision time, the combination of these processes being collectively known as atomic diffusion. This net differential motion leads to he gradual spatial segregation of important chemical elements, and often produces macroscopically observable effects.

Stellar physicists of the beginning of the 20th century recognised and analysed this situation. In his book ``The Internal Constitution of the Stars", \citet{eddington26book} predicted that, due to the radiative acceleration, heavy elements should accumulate at the surface of massive stars, which was not observed at that time. It was  then assumed that macroscopic motions, such as rotational mixing, were strong enough to prevent element segregation. Later on, when the first abundance anomalies were observed in A-type stars, they were first attributed to nuclear effects \citep{fowler65}, although quantitative computations failed to account for the observations \citep[]{michaud70phd, vauclair72, cowley75}. This failure led \citet{michaud70} to propose atomic diffusion as a better explanation. 

From then on, atomic diffusion is stars was studied in depth \citep[][and references therein]{michaud76mcv2, vauclair78v2m, michaud13}. It occurs indeed in all kinds of stars, in a more or less important way according to the stellar situation. First approximate computations of radiative accelerations were done by \citet{michaud76mcv2} who found that they are smaller than gravity inside the Sun, and become of the same order for stars of about 1.2 M$_{\odot}$. For larger masses they are preponderant and become really important in A stars. The resulting abundance variations depend essentially on the competition between radiative levitation and gravitational settling, and of their interaction with macroscopic motions.

When a surface convective zone is present, atomic diffusion  can nevertheless occur in the radiative zone below, which then communicates to the surface any resulting  modification of the chemical composition near the radiative--convective interface. In cool stars like the Sun, the convective zone is deep enough for the atomic diffusion time scale below the radiative--convective interface to be significantly larger than the stellar age. However, atomic diffusion can still cause variations of chemical composition of the order of ten percent compared with models without diffusion \citep[]{aller60, turcotte98}. A revolution in this respect occurred with the avent of helioseismology, which proved that the settling of helium and heavier elements {\it must} be taken into account in the Sun in order to properly account for observations \citep[]{bahcall95, gough96, richard96}.

Renewed attention was recently given to the fact that atomic diffusion can induce significant element accumulation in some layers inside the stars. Although not visible at the surface, such accumulation regions can nevertheless modify the stellar structure \citep[]{richard01}. They are caused by the fact that the ionisation state of each element varies with temperature and density, and hence depth. This effect induces large variations in the contribution of each element to the average opacity, and modifies the radiative acceleration on this element, which is sometimes larger than gravity, sometimes smaller.  For example, when an element is in the noble gas configuration, its contribution to the opacity is small and it undergoes only a small radiative acceleration. The opposite is true when it is in the hydrogenic state. 

The most important element in this respect is iron, which may become the {\it main} contributor to the overall opacity when it is in a ionization state with a very large number of absorption lines. When this is the case, iron ions become subject to a large radiative acceleration which overcomes gravity and pushes them upwards. Above those layers the ionisation state changes and the radiative acceleration drops sharply again. The net result is that iron accumulates at the precise location where its intrinsic opacity is the highest. The increase in iron content then further increases the local opacity of the gas. \citet{richard01} have shown that the induced opacity increase may be large enough to create an extra convective zone, which modifies the mixing of the chemical elements inside the star and may have important consequences for its chemical composition. Such iron accumulation can also induce in some cases an extra $\kappa$-mechanism which may trigger stellar oscillations \citep[]{charpinet97}, as described in Section \ref{sec:pulsations}.

\subsection{Theory vs. observations: the necessary interplay between atomic diffusion and fingering convection}

Generally speaking, classical computations of atomic diffusion in A and F type stars show that, when no extra mixing is taken into account, most heavy elements accumulate near the stellar surface, except some elements like calcium and scandium which may become depleted. However the computed abundance variations are typically much larger than observed, and models can only be reconciled with observations provided extra macroscopic motions are added to mitigate the effects of element segregation. Many studies have investigated atomic diffusion coupled with rotational mixing, micro turbulence, mass loss, or other parametrized macroscopic motions to try and account for observations (e.g. \citet{vick08}). 

In these computations however, the basic instabilities directly induced by the element accumulations themselves were ignored. They should nevertheless be included, since they are expected to occur in any case. The fact that the observed overabundances are not as large as expected from previous computations may simply be related to the self-regulating process which occurs when elements which move up by atomic diffusion are mixed backwards by the induced instabilities. Taking all relevant processes into account will likely change our current understanding of element accumulation in stars.

Indeed, the local abundance increase of important elements like iron can lead to inversions of mean molecular weight gradient, inducing a double-diffusive instability traditionally called ``thermohaline convection" as in the ocean, now preferentially referred to as ``fingering convection" after \citet{Traxleral2011}. Fingering (thermohaline) convection is now recognized as a major mixing process in stellar interiors and has already been studied in several other contexts, including the accretion of heavy elements onto a star, which may be due to planetary material \citep{vauclair2004mfa, garaud11, theado12tv} or to an evolved companion in a binary system \citep{stancliffe07,thompson08}. 

\citet{theado09} were the first to discuss and study the effect of fingering convection on the accumulation of heavy-elements induced by atomic diffusion, focusing in particular on the case of iron. They found that the gradual accumulation process was attenuated by fingering instabilities, but not completely suppressed. Crucially, they also discovered that the helium settling which occurs simultaneously with the heavy element accumulation can stabilize the global mean-molecular weight gradient ($\mu$-gradient hereafter), and thus preserve some of the iron accumulation. They also found that the extra convective zone discussed in \citet{richard01} still persists in some cases, and may have important implications for the element abundances and seismic behavior of several types of stars. 

In these computations, however, the original prescription given by \citet{kippenhahn80} for the mixing coefficient associated with fingering convection was used. Since then, much progress has been made to better quantify mixing by this specific instability, using two-dimensional numerical simulations \citep{denissenkov2010}, and three-dimensional (3D) numerical simulations \citep{Traxleral2011}. The influence of the radiative acceleration on the mixing mechanism itself was also studied by \citet{vauclair12vt}. 

The most recent work on the subject is that of \citet{Brownal2013}, who used 3D numerical simulations combined with theoretical stability analyses to propose and test a new prescription for mixing by fingering convection, that corrects previous inconsistencies of the models of \citet{denissenkov2010} and \citet{Traxleral2011}, and contains no remaining free parameters. In what follows we shall use this latest prescription to model the combined effects of atomic diffusion and mixing by fingering convection on the accumulation of iron in stellar interiors, using both 3D simulations and 1D stellar evolution calculations. Before describing our theoretical approach in more detail, however, we first take a step back and discuss in which ways a better understanding of these combined processes will help us understand observations, focusing in particular on the case of stellar pulsations in the era of {\it Kepler} asteroseimology.

\subsection{Importance for chemical abundances and stellar pulsations} 
\label{sec:pulsations}

A better understanding of atomic diffusion is crucial to improving our understanding of pulsating stars. Indeed several types of stellar pulsators are clearly affected by atomic diffusion and its macroscopic consequences, so that quantitative predictions and comparisons with observations will be different when including the effect of fingering convection. Let us review here a few examples \citep[see also a more detailed discussion in][]{theado09}.

The SPB (slowly pulsating B stars) and the $\beta$ Cephei stars are main sequence B type pulsators. The SPB stars show high-order g-modes, with periods of order 0.5 to 5 days. The $\beta$ Cephei stars show low order p and/or g-modes with periods of 2 to 8 hours. These oscillations are generally thought to be caused by the $\kappa$-mechanism taking place in the metal opacity bump region \citep{cox92, kiriakidis92, moskalik92}, but the mode excitation remains difficult to explain in some of these stars. An increase in the abundance of iron in this region may help resolve this difficulty \citep{pamyatnykh04, miglio07}. It is thus important to compute improved metal abundance profiles in these stars, which properly take into account the effects of fingering instabilities.

The $\gamma$ Doradus stars are A-F main sequence pulsators. They show high-order g-modes with periods between 0.35 to 3 days, thought to be driven by a flux-blocking mechanism at the base of their convective envelope \citep{guzik00,dupret05}. However this process may work only if the convection zone is deeper than given in standard models. Here again, including atomic diffusion and fingering convection could have important consequences on the stellar structure, and could potentially reconcile models with observations.

Finally the subdwarf B stars (sdB stars) are evolved, compact objects which lie on the extended horizontal branch. The hottest ones present rapid oscillations due to low-order, low-degree p-modes, with period of 80 to 600 sec. The coolest ones oscillate with periods of 2000 to 9000 sec, due to high-order, low-degree g-modes. Both types of pulsators are thought to be driven by the $\kappa$-mechanism acting in the iron-peak element opacity bump \citep{charpinet09}, and will therefore be affected by the induced macroscopic motions. A better understanding of fingering convection may help interpret the observations more precisely.

In short, the new detailed seismic observations of the {\it Kepler} satellite \citep[and in the future the Plato experiment, see][]{rauer13}, reveal a great need for improved models of atomic diffusion that include the macroscopic transport induced by fingering instabilities. Our goal is to study the fingering convection induced by local element accumulation (here iron) inside stars, and deduce the possible regimes that may arise when considering atomic diffusion flux and fingering-induced mixing simultaneously. 

\subsection{Our theoretical approach to studying the interplay between atomic diffusion and fingering convection}

In what follows, we first motivate our study by presenting in Section 2 the example of a 1.7 M$_{\odot}$ stellar model where iron accumulates due to the combined effect of gravitational and radiative diffusion, without taking fingering instabilities into account. We show that in this case a strong peak in the iron abundance profile appears, whose shape is close to being Gaussian. 

In order to study the effects of fingering instabilities on the development of this iron layer, we cannot {\it a priori} use the model developed by \citet{Brownal2013} directly. Indeed, the latter was derived and proposed in the much simpler context where atomic diffusion is ignored, and where the local abundance scale height is much greater than the typical vertical scale of the fingering instability. Since the developing iron layer can in some cases be fairly thin, it is not clear whether this assumption remains correct or not. In order to study the validity of the Brown et al. (2013) prescription, we first present in Section 3 to 6 the first 3D hydrodynamic simulations of the heavy element accumulation process, which naturally lead to the emergence of the fingering instability. We first lay out the model equations in Section 3, and discuss the various parameter regimes observed in these simulations, according to the input parameters, in subsequent Sections (4-6). By comparing the results of our 3D simulations with those of a one-dimensional model that uses the prescription of Brown et al. (2013), we find in Section 7 that the latter is a remarkably good model for fingering convection, even in this more complex context. We then implement it, in Section 8, in the Toulouse-Geneva Evolution Code and study the result of iron accumulation and fingering convection for the model star introduced in Section 2. We find that fingering convection alone mostly erases the iron peak, but the latter can be preserved when helium settling is also taken into account, as originally found by \citet{theado09}. Finally, we also find that convection zones can indeed spontaneously emerge from the accumulation of iron, as suggested by \citet{richard01}, and are triggered by the strong increase in the opacity in the iron layer. Implications of our findings are discussed in the conclusion. 

\section{Background stellar model} 

In this section we begin with a simple example and present the results of a stellar model obtained for a 1.7 M$_{\odot}$ star at various ages, to show the formation of the iron peak, when atomic diffusion including radiative levitation and gravitational settling are taken into account, but without introducing fingering convection.

The models were computed using the Toulouse-Geneva Evolution Code (TGEC), which includes atomic diffusion with radiative accelerations computed in a precise way for 21 species, namely 12 elements and their main isotopes: H, $^{3}$He, $^{4}$He, $^{6}$Li, $^{7}$Li, $^{9}$Be, $^{10}$B, $^{12}$C, $^{13}$C, $^{14}$N, $^{15}$N, $^{16}$O, $^{17}$O,$^{18}$O, $^{20}$Ne, $^{22}$Ne, $^{24}$Mg, $^{25}$Mg, $^{26}$Mg, $^{40}$Ca and $^{56}$Fe \citep{theado12talv}. The diffusion computations are based on the Boltzmann equation for a dilute collision-dominated plasma. When the medium is isotropic, the solution of the Boltzmann equation is a Maxwellian distribution function. In stars however, structural gradients (temperature, pressure, density, etc.) lead to small deviations from the Maxwellian distribution, which are specific to each species. Solutions of the Boltzmann equation are then obtained in terms of convergent series representing successive approximations to the true distribution function \citep{chapman70}. The computations lead to a statistical ``diffusion" or ``drift" velocity $w_d$ of the element with respect to the main component of the plasma. The abundance variations of all the elements are computed simultaneously, with the use of the mass conservation equation. The gravitational and thermal diffusion coefficients used in the code are those derived by \citet{paquette86}. 

OPCD v3.3 codes and data \citep{seaton05} are used to compute self-consistent Rosseland opacities at each time step to take the variations of the abundances of each element into account. Using these consistent opacities the code computes radiative accelerations on C, N, O, Ne, Mg, Ca, and Fe using the improved semi-analytical prescription proposed by \citet{alecian04}. The computations of the total radiative accelerations on all these elements require computing the relative populations of each ion of each element. This is included in the numerical routines of TGEC.

The equation of state used in the code is the OPAL2001 equation \citep{rogers02}. The nuclear reaction rates are from the NACRE compilation \citep{angulo99}. The mixing length formalism is used for the convective zones with a mixing length parameter of 1.8, as needed to reproduce solar models.

The results are presented for three ages at the beginning of the main-sequence evolution of the star, at 23, 28 and 35 Myrs. In Figure 1 we show the variation with radius of the radiative acceleration on iron and of gravity at 35 Myrs. We can see that the radiative acceleration strongly varies with depth, due to the changes in the iron ionisation state. The layer where it is the largest, around $0.97$ $10^{11}$ cm, corresponds to the so-called ``opacity bump" where iron is the main contributor to the average opacity. Above that layer, the radiative acceleration drops dramatically, thereby causing a local iron accumulation around $r_0 = 10^{11}$cm. Note that below, around $0.90$ $10^{11}$ cm, the contrary happens, leading to a local iron depletion.

In Figure 2, we present the formation and evolution of the iron peak for the three ages. The peak forms rapidly and the iron overabundance grows to be quite large. In spite of the original asymmetries due to the density, pressure and temperature stratification inside the star, the peak rapidly acquires a nearly Gaussian shape. Table 1 gives the values of the physical parameters inside the star at the region of the iron peak. 

\begin{figure}
\begin{center}
\includegraphics[width=4.5in]{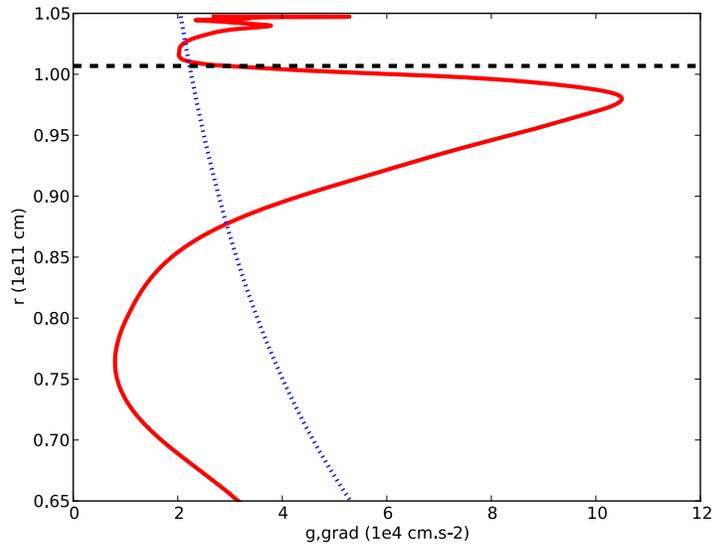}
\caption{Radiative acceleration (solid red line) on iron ions in a 1.7 M$_{\odot}$ star at 35 Myrs without fingering convection, compared with gravity (dotted blue line) as a function of radius near the iron layer. The black dotted line represents the position of the iron peak.}
\label{grad}
\end{center}
\end{figure}

\begin{figure}
\begin{center}
\includegraphics[width=4.5in]{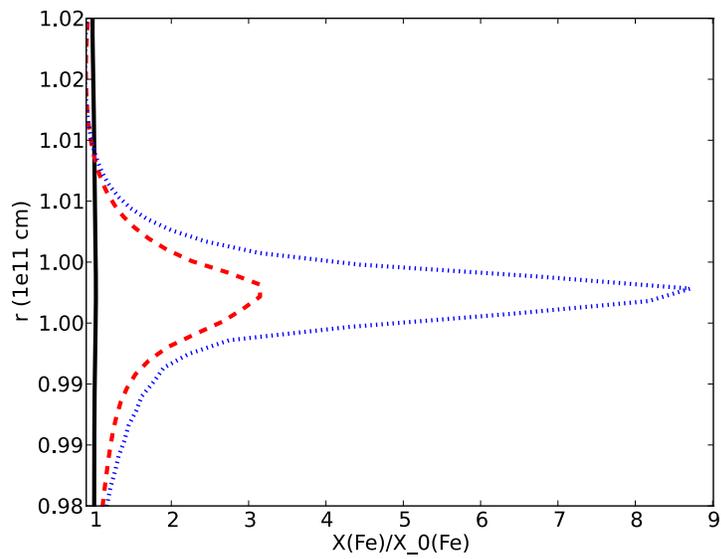}
\caption{Iron abundance profiles in a 1.7 M$_{\odot}$ star at 23Myrs (black solid line), 28Myrs (red dashed line) and 35 Myrs (blue dotted line) without fingering convection. }
\label{Fe-noth}
\end{center}
\end{figure}


\begin{table}
\center
\begin{tabular}{ | c | c | c | c | c | c | c | c | c | c | c | c |}
\hline  $P~(10^8)$ & $T~(10^5)$ & $\rho~(10^{-6})$ & $\nabla_\mathrm{ad}$ & $\nabla_\mathrm{rad}$ & $H_\mathrm{p}~(10^{9})$ & $\log(g)$  & $\kappa$ & $\nu$ & $\kappa_\mathrm{\mu}$ & $\kappa_\mathrm{T}~(10^{12}) $  \\
\hline   1.21 & 1.62 & 5.34 & 0.397 & 0.277 & 1.01 & 4.348 & 41 & 3370 & 27 & 2.4    \\
\hline
 \end{tabular}
 \caption{Physical parameters in the iron peak region of a 1.7 M$_{\odot}$ model at 35 Myrs. The table features the local pressure, temperature, and density, the adiabatic and radiative gradients, the pressure scale height, gravity, the opacity, the kinematic viscosity (sum of the molecular and radiative viscosities), the molecular diffusivity and the thermal diffusivity. All units are in cgs.}  
\end{table}

These results have been obtained in stellar models computed with pure atomic diffusion, without any additional mixing in the iron layer. We know however that such an iron accumulation is not stable, due to the induced inversion of the mean molecular weight gradient. The inversion causes fingering convection which has to be added in the computations. As discussed above, the most recent model of fingering convection is that of Brown et al. (2013), but the latter cannot {\it a priori} be applied when atomic diffusion, and strongly varying background compositional gradients, are present. In what follows, we therefore run and study 3D simulations of the iron layer formation and consequent evolution, and test the validity of the Brown et al. (2013) model against these simulations, before returning to the problem of stellar evolution in Section 8. 

\section{A simplified model of the iron layer}

In order to study the fingering dynamics of the iron layer, we must move away from standard stellar evolution codes (which assume hydrostatic equilibrium), and use the complete Navier-Stokes equations to describe the problem. However, following the evolution of a fully resolved 3D fingering field while at the same time modeling radiative transfer processes in detail is numerical impossible at the present time. For this reason, we first propose a simplified model of the iron layer that nevertheless still captures most of the basic physics of the problem. 
 
Since the expected size of fingering structures is much smaller than the star \citep{schmitt1983csf}, we cannot model them in a whole-star simulation. Instead, we consider a small Cartesian domain in the vicinity of $r_0$, that is tall enough to include the whole iron layer and whose horizontal extent is sufficiently large to include a representative number of expected fingering structures. The $z$-direction is aligned with gravity (with $z = r - r_0$), the $x$-direction is aligned with the azimuth, and the $y$-direction with latitude. We shall however ignore any effect of rotation, so that the local gravity is the only source of anisotropy in the system.

Since the layer itself is reasonably thin compared with a typical pressure/temperature scaleheight, we use the Boussinesq approximation \citep{spiegelveronis1960} to model its dynamics. This approximation assumes that the total density and temperature fields can be written as 
\begin{eqnarray}
&& \rho_{\rm tot}(x,y,z,t) = \rho_0 + z \bar \rho_z + \rho(x,y,z,t) \mbox{   , } \nonumber \\
&& T_{\rm tot}(x,y,z,t) = T_0 + z \bar T_z + T(x,y,z,t)  \mbox{   , } 
\end{eqnarray}
where $\rho_0$ and $T_0$ are the typical mean density and temperature near $r_0$ prior to the accumulation of iron, and where perturbations around these mean values are small and expressed as a linear function of $z$ only plus a space- and time-dependent function. The latter, $\rho$ and $T$, are of course related to one another via the equation of state, but also depend on the local density of iron, called $C$ hereafter:
\begin{equation}
\frac{\rho}{\rho_0} = - \alpha T + \beta C \mbox{   , } 
\end{equation}
where  $\alpha$ and $\beta$ are determined by linearizing the full equation of state around $\rho_0$. This yields
\begin{equation}
\alpha = - \frac{1}{\rho_0} \left( \frac{\partial \rho}{\partial T} \right)_{p = {\rm constant},C=0} \quad \mbox{   and   } \beta = \frac{1}{\rho_0}  \mbox{   . } 
\end{equation}
Dimensionally, $C$ has units of g/cm$^3$. 

The local background temperature gradient $\bar T_z$ and adiabatic temperature gradient $\bar T^{\rm ad}_z$ are assumed to be both constant within the modeled domain, and so is gravity. In what follows, we also need to introduce the standard microscopic diffusivities (viscosity $\nu$, thermal diffusivity $\kappa_T$ and iron diffusivity $\kappa_C$). The latter are assumed to be constant, and are related to the atomic diffusion rates discussed in the introduction. However, this standard diffusion proceeds down-gradient, by contrast with atomic diffusion (gravitational settling, radiative levitation) which does not necessarily do so. Note that the assumption of a constant thermal diffusivity is possibly the weakest component of this simplified model, since it prevents any possibility of triggering convection by changes in the local opacity. However, this assumption is necessary for numerical computations since the latter are much more difficult to implement and run efficiently when the diffusivities are nonlinear functions of temperature and composition. 

Within these assumptions, the governing equations for the system are then \citep{spiegelveronis1960}:
\begin{eqnarray}
&& \frac{\partial \vec{u}}{\partial t} + \vec{u}\cdot\nabla \vec{u} = -\frac{1}{\rho_0} (\nabla p - \rho \vec{g}) + \nu \nabla^2 \vec{u}  \mbox{   , }   \nonumber \\
&& \frac{\partial T}{\partial t} + \vec{u}\cdot\nabla T + w \left(\bar{T}_z- \bar{T}^{\rm ad}_z \right) = \kappa_T \nabla^2 T  \mbox{   , }  \nonumber \\
&& \frac{\partial C}{\partial t} + \vec{u}\cdot\nabla C + \frac{\partial}{\partial z} (w_d(z) C) = \kappa_C \nabla^2 C  \mbox{   , }  \nonumber \\
&& \nabla \cdot \vec{u} = 0  \mbox{   , } 
\end{eqnarray}
where $\vec{u} = (u,v,w)$ is the velocity field, $p$ is the pressure, and $w_d(z)$ is the iron drift velocity caused by the combination of gravitational settling and radiative levitation (see below).

These equations are similar to the ones used to describe fingering convection by \citet{Brownal2013}, except for the use of $C$ instead of the mean molecular weight $\mu$, and the treatment of the $C$-equation itself. Indeed, in
 \citet{Brownal2013} fingering convection is driven by an assumed constant background $\mu$-gradient. In our particular problem, on the other hand, the initial iron density is very small and constant (assuming that the star is well mixed on the ZAMS), but changes with time first as a result of the competing effects of gravitational settling and radiative levitation, and later from fingering convection. As a result, it is preferable to assume that there is no background iron density gradient to begin with, and merely follow the evolution of the total iron density profile $C$. 

However we must now add the effects of settling and levitation. This is done through added ``drift" velocity term $w_d$, which is a function of the vertical position $z$. For simplicity, instead of using the drift velocity profile computed from the model presented in Section 2, we use a simple analytical function to model $w_d(z)$. Since iron is settling from above, and levitating from below, we model $w_d$ as a monotonically decreasing function of $z$ that vanishes exactly at $z = 0$. Except when specifically mentioned, we shall take $w_d(z)$ to be a linear function of $z$, an approximation that can simply be viewed as a Taylor-expansion of the true function $w_d(z)$ near $z=0$. As long as the element accumulation layer is quite thin compared with the typical lengthscale over which $w_d(z)$ naturally varies, this is quite a good approximation.  
 
 These equations are then non-dimensionalized with the following length and time scales, as described in \citet{Traxleral2011} for instance:
\begin{eqnarray}
&& d = \left[l\right] = \left(\frac{\kappa_T \nu} {g \alpha (\bar{T}_z-\bar T^{\rm ad}_z) }\right)^{1/4}  \mbox{   , }  \nonumber \\
&& \left[t\right] = \frac{d^2}{\kappa_T} , \quad \, \left[u\right] = \frac{\kappa_T}{d}  \mbox{   , }  \nonumber \\
&& \left[T\right] = (\bar{T}_z-\bar T^{\rm ad}_z) d , \quad \, \left[C\right] = \frac{\alpha }{\beta} (\bar{T}_z-\bar T^{\rm ad}_z) d   \mbox{   . } 
\end{eqnarray}
In this non-dimensionalization, the unit lengthscale $d$ is related to the typical horizontal length scale of a finger (to be specific, a finger in this parameter regime is typically of the order of 10$d$), and the unit timescale is the thermal diffusion time scale across $d$. The governing equations can then be re-written as follows: 
\begin{eqnarray}
&& \frac{1}{\rm{Pr}}\left(\frac{\partial \vec{u}}{\partial t} + \vec{u}\cdot\nabla \vec{u}\right) = -\nabla p + (T-C)\hat{\bf e}_k + \nabla^2 \vec{u}  \mbox{   , }  \nonumber \\
&& \frac{\partial T}{\partial t} + \vec{u}\cdot\nabla T + w  = \nabla^2 T  \mbox{   , } \nonumber \\
&& \frac{\partial C}{\partial t} + \vec{u}\cdot\nabla C - s \frac{\partial}{\partial z} (z  C) = \tau \nabla^2 C  \mbox{   , } \nonumber \\
&& \nabla \cdot \vec{u} = 0\mbox{   , } 
\label{eq:nondim}
\end{eqnarray}
where all quantities are now non-dimensional. This set of equations contains three non-dimensional parameters: ${\rm Pr}$, $\tau$ and $s$. $\rm{Pr}  = \nu / \kappa_T$ is the usually-defined Prandtl number, and $\tau = \kappa_C / \kappa_T$ is the so-called diffusivity ratio. Finally, the non-dimensional drift velocity is expressed as $w_d(z)d/\kappa_T  = - sz$, where $s$ can be interpreted as a ``pinching rate", and if expressed dimensionally, would indeed have the dimensions of one over time. 

An additional non-dimensional parameter arises from the choice of initial conditions. Assuming that the total mass of iron in the accumulation layer does not vary much with time (i.e. the density profile merely changes shape while conserving the total iron content), we can define $\Sigma_0$ as mean surface density of iron between the bottom ($z_b$) and the top ($z_t$) of the layer. This quantity is constant, and constrains the evolution of $C$ via 
\begin{equation}
\Sigma_0 = \frac{1}{L_xL_y} \int_{x=0}^{L_x}\int_{y=0}^{L_y} \int_{z=z_b}^{z_t} C dxdydz  \mbox{   . } 
\end{equation}
Going back to dimensional, stellar quantities, the non-dimensional surface density $\Sigma_0$ is related to the dimensional quantity $M_0$, the total mass of iron in the layer, via
\begin{equation}
M_0 \simeq \frac{\alpha}{\beta} (\bar T_z - \bar T^{\rm ad}_z)  d^2 4 \pi r_0^2 \Sigma_0    \mbox{   . } 
\end{equation}
This is only an approximation, whose quality depends on the geometrical ratio of the thickness of the iron layer to $r_0$. However, since $M_0$ can easily be estimated from a stellar evolution code, this equation provides a fairly simple way of estimating plausible values of $\Sigma_0$ in actual stellar conditions. 

For the stellar model presented in the previous section, we can deduce typical values of the various dimensional and non-dimensional quantities of importance for fingering convection. The latter are presented in Table 2. 
\begin{table}[h]
\center
\begin{tabular}{ | c | c |}
\hline  $d$  & $ \sim 3 \times 10^5$ cm \\
\hline  $M_0$ & $ \sim 4 \times 10^{26}$ g \\
\hline Pr & $ \sim 1.5 \times 10^{-9}$ \\ 
\hline $\tau $ &  $ \sim 10^{-11} $ \\ 
\hline $ s $ & $\sim  6 \times 10^{-17} $ \\
\hline $\Sigma_0$ & $\sim 10^8$ \\
\hline
 \end{tabular}
 \caption{Physical parameters in the iron peak region of a 1.7 M$_{\odot}$ model at 35 Myrs. The first two entries are the typical finger width, $d$ and total mass of iron in the accumulation layer, $M_0$, in cgs units. The next entries are all non-dimensional.}  
\end{table}

\section{Evolution of the background profiles prior to the onset of instability}
\label{sec:laminar}

We begin our investigation by considering the evolution of the iron density $C$ assuming that there is no fluid motion, that is, prior to the onset of any form of instability. Thus, we seek solutions of the one-dimensional advection-diffusion equation for the laminar profile $C_l(z,t)$: 
\begin{equation}
\frac{\partial C_l}{\partial t} - s \frac{\partial}{\partial z} (z C_l) = \tau \frac{\partial^2 C_l}{\partial z^2}  \mbox{   . } 
\label{eq:ad_difflaminar}
\end{equation}
It is easy to show that a possible analytical solution to this equation\footnote{To be specific, this solution is only strictly valid when the initial condition for $C$ is also a Gaussian. However, it can be verified numerically that this time-dependent Gaussian is an {\it attracting} solution of the true profile, that is, exact solutions to the problem for arbitrary initial conditions rapidly converge to a Gaussian (see Figure \ref{fig:S=0.5Scrit} for instance).} is a time-dependent Gaussian of the form:
\begin{equation}
C_l(z,t) = \frac{\Sigma_0}{\sqrt{2 \pi }\Delta(t)} e^{- \frac{z^2}{2 \Delta^2(t)}}  \mbox{   , } 
\label{eq:laminarsolution}
\end{equation}
where $\Delta(t)$ is the width of the Gaussian at time $t$. Plugging (\ref{eq:laminarsolution}) into (\ref{eq:ad_difflaminar}), we can solve for $\Delta(t)$:
\begin{equation}
\Delta^2(t) = \frac{\tau + b e^{-2 s t}}{s}  \mbox{   , } 
\label{eq:laminargauss_width}
\end{equation}
where $b$ depends on the initial conditions applied to $C_l$. Note that $\Delta(t) \rightarrow \sqrt{ \tau/s}$ as $t \rightarrow +\infty$. Therefore, in absence of instabilities the ultimate laminar steady state density profile is:
\begin{equation}
C_\infty (z) = \frac{\Sigma_0}{\sqrt{2 \pi} \Delta_\infty} e^{-\frac{z^2}{2 \Delta^2_\infty}} \quad \mbox{ where } \Delta_\infty = \sqrt{\frac{\tau}{s}}  \mbox{   . } 
\label{eq:laminarsteady}
\end{equation}
The dependence of $\Delta_\infty$ on $s$ and $\tau$ is not surprising: $\Delta_\infty$ is smaller if (1) the ``pinching rate" $s$ is larger or (2) if the iron diffusivity, represented by $\tau$ in a non-dimensional sense, is smaller. The rate at which the element layer contracts to the laminar steady-state solution is $s$. 

For the typical stellar values of $\tau$ and $s$ given in Table 2, we find that $\Delta_\infty \sim 400$, which in dimensional terms becomes $\Delta_\infty = 400d \sim 1.2 \times 10^8$cm. Note that this is significantly larger than the vertical scale of basic fingering structures, which are typically of the order of $10d$.

\section{The onset of fingering instability}
\label{sec:onset}

As iron accumulates near $z=0$, an inverse $\mu-$gradient is created and gradually gains in amplitude. As first discussed by \citet{theado09} and reviewed in the introduction, this can trigger fingering instabilities. 
To determine at which point in time the onset of fingering will occur, one should perform a formal stability analysis of the evolving $C_l(z,t)$ profile. However, studying the stability of a time-dependent system is never particularly easy. A commonly used approximation to simplify the problem is the so-called ``frozen-in" limit, where one studies instead the stability of successive profiles which are each assumed to be in a steady state. This usually yields a satisfactory estimate of the properties of the onset of instability, a statement we further verify in the next Section. 

Henceforth, we consider the stability of the laminar iron density profile $C_l(z,t_0)$ (given by equation \ref{eq:laminarsolution}) around a selected time $t_0$. 
The most unstable point of this density profile is the one for which $dC_l/dz$ is the largest (i.e. the point where the inverse $\mu$-gradient is the strongest). This happens at the lower inflection point of the Gaussian, $z_i = - \Delta_l(t_0)$. At this point, 
\begin{equation}
\frac{dC_l}{dz} = \frac{\Sigma_0}{\sqrt{2 \pi e}\Delta^2_l(t_0)}  \mbox{   . }
\end{equation}

\citet{bainesgill1969} showed that the stability of a system with {\it constant} background temperature and compositional gradients, and in the absence of settling/levitation, is uniquely determined by ${\rm Pr}$ and $\tau$ as well as the value of the density ratio $R_0$, defined as
\begin{equation}
R_0 = \frac{\alpha (\bar T_z - T^{\rm ad}_z) }{\beta \bar C_z} = \frac{1}{dC_l/dz} \mbox{   , }
\label{eq:densityratio}
\end{equation}
where all quantities in the first right-hand-side of that expression are dimensional, and all quantities in the second right-hand-side of that expression are non-dimensional. \citet{bainesgill1969} also found that the system is stable to fingering convection, unless $1 < R_0 < \tau^{-1}$. 
While our particular problem does have a constant background temperature gradient, the compositional gradient clearly varies with $z$, so their result cannot formally be applied here. Furthermore, the
presence of a non-zero drift velocity $w_d(z)$ could {\it a priori} affect the linear stability of the problem. 

However, $w_d = -sz$ is so small (see Table 2) that it is unlikely to have a significant influence on the onset of fingering instability, and on the characteristics of the most unstable modes. This, again, is verified in the following section. Furthermore, if we assume that the instability first develops in the vicinity of the most unstable point only, and that the typical mode length scale near onset is much smaller than the 
scale height of the iron density profile (which was verified in the previous Section), then we can get an estimate for the stability of the system simply by considering the value of the {\it local} density ratio near that point: 
\begin{equation}
R(t_0) \equiv R(z_i(t_0),t_0) =  \left. \frac{1}{dC_l/dz} \right|_{(z_i(t_0),t_0)}= \frac{\sqrt{2 \pi e}\Delta^2_l(t_0)}{\Sigma_0} \mbox{   , }
\end{equation}
where all quantities are now expressed in their non-dimensional form. At early times, the Gaussian is presumably very thick ($\Delta_l(t_0)$ is very large) and so $R(t_0)$ is usually much larger than $1/\tau$. However, as iron accumulates near $z=0$, $\Delta_l(t_0)$ gradually decreases (as expressed in equation (\ref{eq:laminargauss_width})), and so does $R(t_0)$. Eventually, $R(t_0)$ becomes equal to $1/\tau$, which then triggers the onset of the fingering instability. This happens when 
\begin{equation}
\Delta(t_0)^2 =  \frac{\Sigma_0}{\tau \sqrt{2 \pi e}}  \equiv \Delta^2_i  \mbox{   . }
\label{first_instab}
\end {equation}
which defines the thickness of the layer at onset, $\Delta_i$.

We then see that whether the iron layer ever becomes unstable or not depends on whether $\Delta_i > \Delta_\infty$ or $\Delta_i < \Delta_\infty$. Indeed, if $\Delta_\infty > \Delta_i$, the evolving density profile would settle to a laminar steady state {\it before} reaching the width at which fingering instabilities appear. In the opposite scenario, fingering convection is expected to set in as soon as the layer thickness drops below $\Delta_i$. Interestingly, we find that whether a system belongs to the first or the second category depends only on $\Sigma_0$, $\tau$ and $s$. To see this, note that we can define a critical surface density $\Sigma_{\rm crit}$ above which fingering convection is expected to occur (i.e. by setting $\Delta_\infty = \Delta_i$ and solving for $\Sigma$). We get
\begin{equation}
\Sigma_{\rm crit} = \frac{\tau^2 \sqrt{2 \pi e}}{s} \mbox{   . }
\label{eq:Scrit}
\end{equation}
If $\Sigma_0 < \Sigma_{\rm crit}$ the system ultimately settles into a laminar steady state, while fingering convection prevents this from happening if $\Sigma_0 > \Sigma_{\rm crit}$. 

\begin{figure}[h]
\begin{center}
\includegraphics[width=0.5\textwidth]{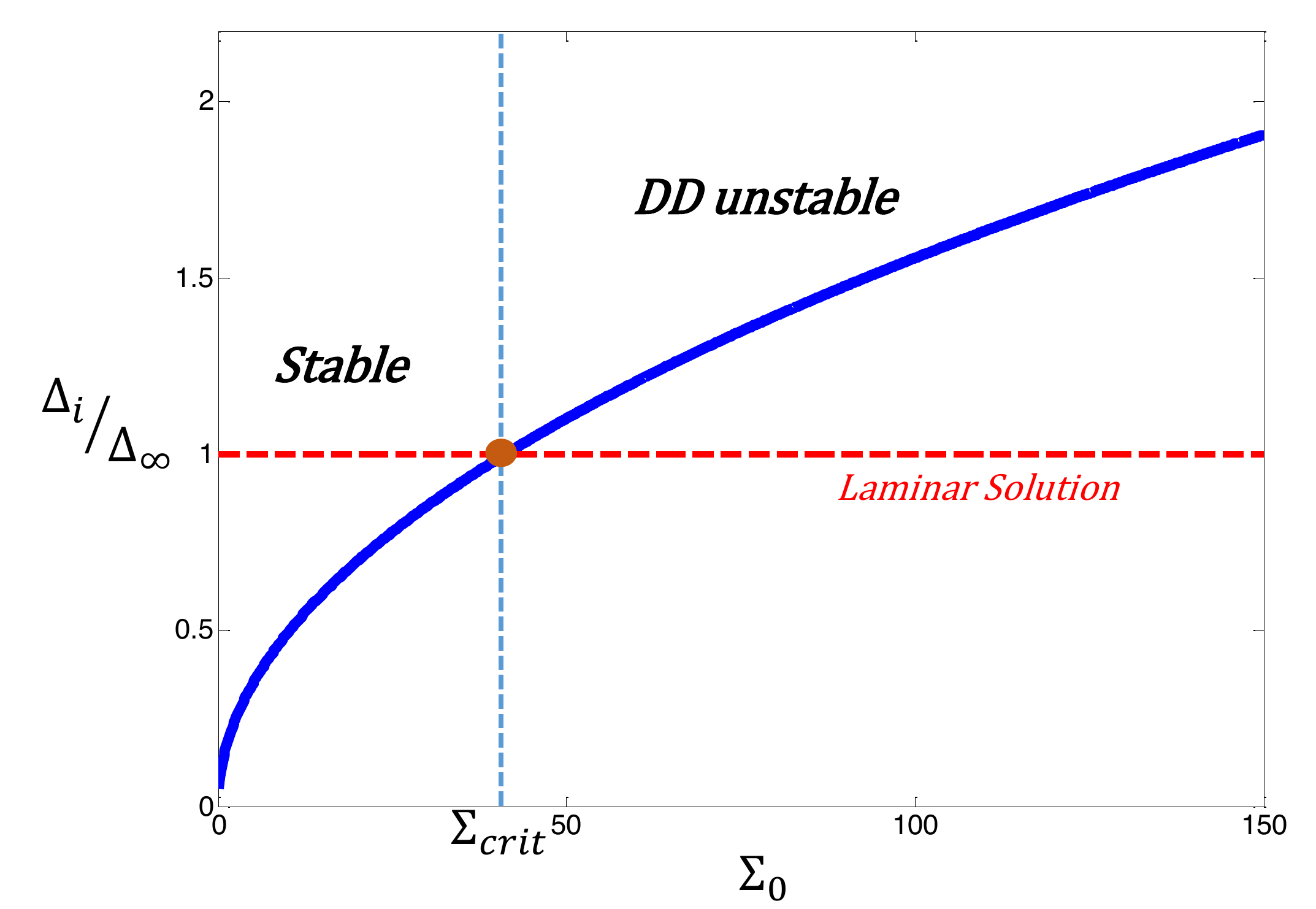}
\caption{Diagram of stable regions and double-diffusively unstable regions with respect to input mass $\Sigma_0$. The system is stable if $\Delta_i/\Delta_{\infty} \leq 1$ ($\Sigma_0 < \Sigma_{\rm crit}$) and unstable if $\Delta_i/\Delta_{\infty} \ge 1$ ($\Sigma_0 > \Sigma_{\rm crit}$). In this figure, we used ${\rm Pr } = \tau =0.1$, and $s = 0.001$, in which case $\Sigma_{\rm crit} \simeq 41$ (see equation \ref{eq:Scrit}). }
\label{fig:instab_region}
\end{center}
\end{figure}

The various regimes, for a particular case with ${\rm Pr } = \tau =0.1$, and $s = 0.001$ (a set of parameters that will be used in the numerical simulations presented in Section \ref{sec:num}), are shown in Figure \ref{fig:instab_region}. For the stellar parameters given in Table 2, on the other hand, $\Sigma_{\rm crit} \ll 1$ while $\Sigma_0 \gg 1$. Hence fingering convection is {\it always} expected to play an important role in the dynamics of accumulating iron layers. The subsequent evolution of the iron density profiles for systems where $\Sigma_0 > \Sigma_{\rm crit}$ is analyzed in the next section.

\section{Numerical simulations of the layer evolution during and past the onset of instability}
\label{sec:num}

In order to study the effects of the fingering instability (and other instabilities) past onset, it is best to use 3D numerical simulations. To do so, we use a code very similar to the one first presented and used by \citet{Traxleral2011b}, modified as described below for our current purpose.

\subsection{Set-up of the 3D code}

As in \citet{Traxleral2011} \citep[see also][]{Traxleral2011b,Brownal2013}, our code solves the set of equations (\ref{eq:nondim}) in a triply-periodic domain. This may seem at first like an odd choice, since the $C-$profile is not expected to be triply-periodic in this particular problem. However, by choosing the vertical extent of the domain to be large enough and by placing the accumulation layer in the middle of that domain, we can ensure that (1) the regions far above and far below the iron layer are at rest (i.e. without any fluid motion), and (2) the iron density is vanishingly small near the top and bottom boundaries. This guarantees that boundary conditions have as little influence as possible on the dynamics near the accumulation layer. 

An additional modification to the code is the new drift velocity term in the iron density equation. While this modification is numerically trivial, note that we also have to chose $w_d(z)$ to be periodic in the vertical direction. Specific choices of $w_d(z)$ are discussed below. The implementation of the new term was satisfactorily tested against analytical and semi-analytical solutions of the problem (see below for some examples). 

In all simulations presented in the following sections, the temperature perturbations were initialized with random low-amplitude noise. The initial iron density profile was taken to be a Gaussian of the form:
\begin{equation}
C(x,y,z,0) = \frac{\Sigma_0}{\sqrt{2 \pi} \Delta_{\rm init}} e^{-z^2 / 2 \Delta_{\rm init}^2}  + \mbox{noise}  \mbox{   , } 
\end{equation}
where $\Delta_{\rm init}$ was chosen to be strictly greater than $\Delta_i(\Sigma_0)$ (see equation \ref{first_instab}), so that the iron layer is initially stable. By doing this we can study the onset of instability, and determine whether it occurs as predicted. To save computational time, however, $\Delta_{\rm init}$ was typically taken to be fairly close to $\Delta_i(\Sigma_0)$, usually between $1.1\Delta_i(\Sigma_0)$ and $2 \Delta_i(\Sigma_0)$. Note that $\Delta_i(\Sigma_0)$ does depend on the input surface density, so that $\Delta_{\rm init}$ must be proportionally larger for larger $\Sigma_0$.

The computational domain size selected was, for each simulation, a tradeoff between computational feasibility and desired result. 
The typical width of a double-diffusive finger pair (one going up and one going down) at this parameter regime being $7d-10 d$, the width of the domain (in the $x$-direction) was chosen in all simulations to be $L_x = 100d$ to allow at least 20 fingers to develop in the system. As shown by Garaud \& Brummell (in preparation), the dynamics of fingering convection in 2D numerical simulations at low Prandtl number show pathological behavior through the spurious development of shear layers that are not seen in 3D simulations at the same parameter regime. To avoid them, we have to run 3D simulations. However, Garaud \& Brummell also showed that a computational domain does not have to be very thick to properly account for the full 3D dynamics of fingering convection, at least in parameter regimes where no further instabilities are expected. Following their results, we choose the thickness of our domain to be $L_y = 15d$, unless otherwise mentioned.  

\begin{figure}[h]
\centerline{\includegraphics[width=0.5\textwidth]{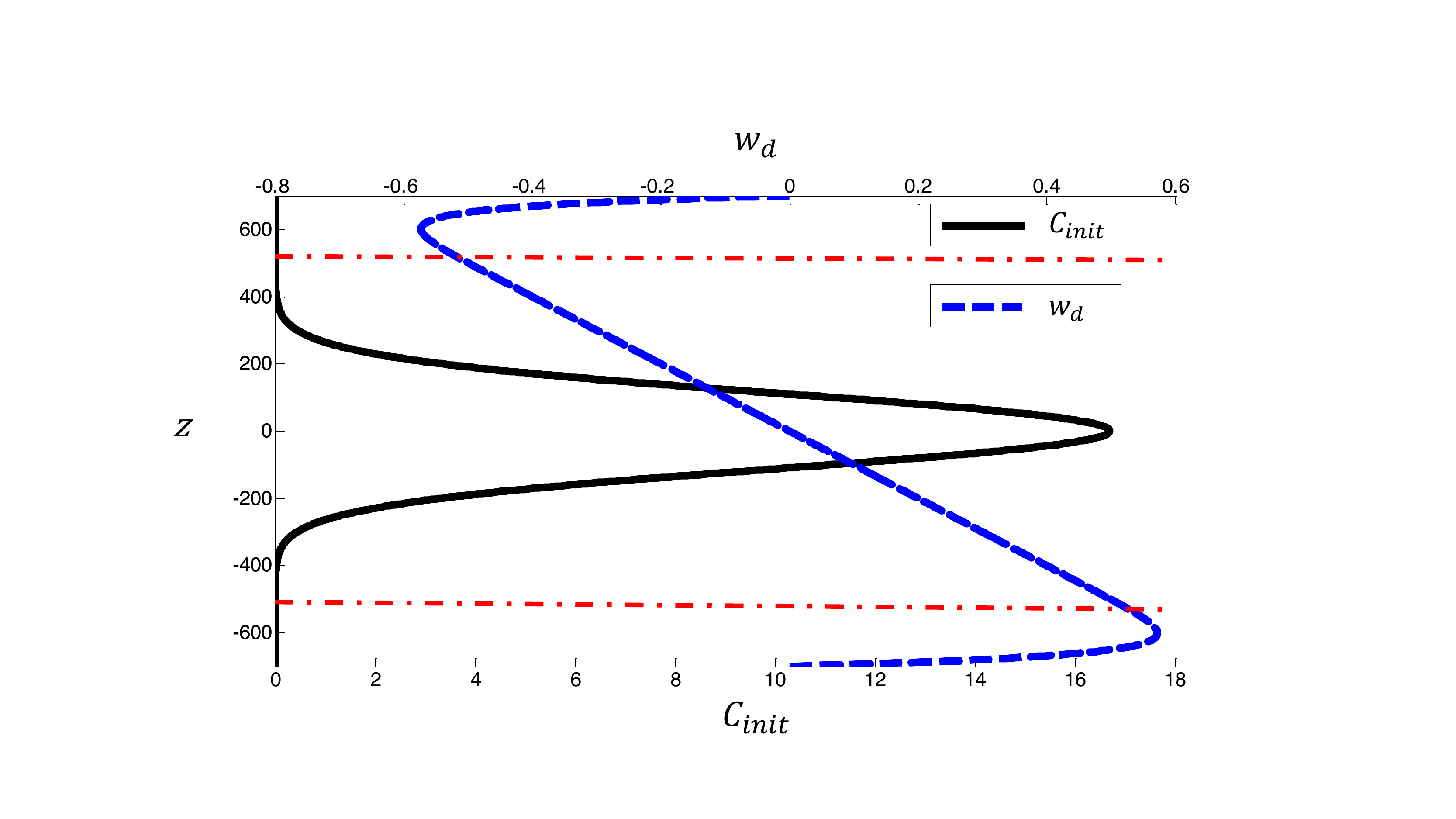}}
\caption{Initial condition (solid black line) and drift velocity profile $w_d(z)$ (blue dashed line) for the following parameters: $\tau = \Pr = 0.1$, $s = 0.001$ and $\Sigma_0 = 100\Sigma_{\rm crit}$. Since $\Delta_i = 100$ at these parameters, we used $\Delta_{\rm init} = 110$. The horizontal red dashed lines indicate the region where the function $w_d(z)$ is reasonably close to being linear. }
\label{fig:forcing}
\end{figure}

The height of the domain ($z$ direction) was varied according to the constraints placed on the problem by the selection of the other input parameters of the simulation (notably, $\Sigma_0$). This selection is constrained by two factors. First, since $\Delta_{\rm init}$ must be larger for larger $\Sigma_0$, and since we require that $C(x,y,z,0)$ must become numerically small near the top and the bottom of the domain, we have to increase $L_z$ in proportion to any increase in $\Delta_{\rm init}$. Secondly, as noted earlier, $w_d(z)$ must be periodic in $z$, but must also vanish near the top and bottom boundaries to avoid any mass flux through the latter. Third, $w_d(z)$ should also be reasonably smooth, and have a reasonably large region near $z=0$ where it is well-approximated by $w_d(z) \sim -sz$. Finally, the bulk of the initial Gaussian profile selected must be able to fit within the region of the domain where $w_d(z)$ is linear, so that the simulations can be compared more directly with the results of Sections \ref{sec:laminar} and \ref{sec:onset}. Based on these statements, we chose to use $w_d(z) = - s z (m^{20} - z^{20})/(m^{20})$, where $m = L_z /2$. This formulation ensures that the region where this function is close to linear encompasses a large fraction of the computational domain, as shown in Figure \ref{fig:forcing}.

Finally, note that it is not computationally feasible to run simulations at stellar values of the parameters $s$, $\rm{Pr}$ and $\tau$. Also, given the very large domain size considered and constraints on the required resolution for very low $\rm{Pr}$ and $\tau$ runs, we cannot reach parameter values as low as \citet{Brownal2013} did. For consistency throughout this work we shall use ${\rm Pr} = \tau  = 0.1$ in all 3D numerical simulations. We also (arbitrarily) set the value of $\Delta_\infty = 10$, so that $s = 10^{-3}$ in all 3D simulations. We then vary the total input mass of iron present via $\Sigma_0$ and study the various parameter regimes that naturally emerge. The role of simulations at non-stellar parameters is thus two-folds: (1) to identify interesting behavior, and (2) to validate the compositional mixing parametrization of \citet{Brownal2013}. This validation will then enable us to apply these simple mixing models to actual stellar evolution codes that use realistic stellar parameters. 

\subsection{Low input mass regime ($\Sigma_0 \sim \Sigma_{\rm crit}$)}

While we have found earlier that the typical values of $\Sigma_0$ expected in real stellar interiors are usually much larger than $\Sigma_{\rm crit}$, it is worth exploring the parameter regimes where $\Sigma_0$ is close to $\Sigma_{\rm crit}$, for completeness, and also as a test of the validity of the code. 


For $\tau = 0.1$ and $s = 0.001$, equation (\ref{eq:Scrit}) shows that $\Sigma_{\rm crit} = 41$. We then examined systems with $\Sigma_0 = 0.5 \Sigma_{\rm crit}$ and $\Sigma_0 = 2 \Sigma_{\rm crit}$ to determine whether (1) the laminar solution indeed behaves as predicted in Section \ref{sec:laminar}, and (2) the onset of fingering convection occurs as predicted in Section \ref{sec:onset}. 

\begin{figure}[h]
\centerline{\includegraphics[width=0.8\textwidth]{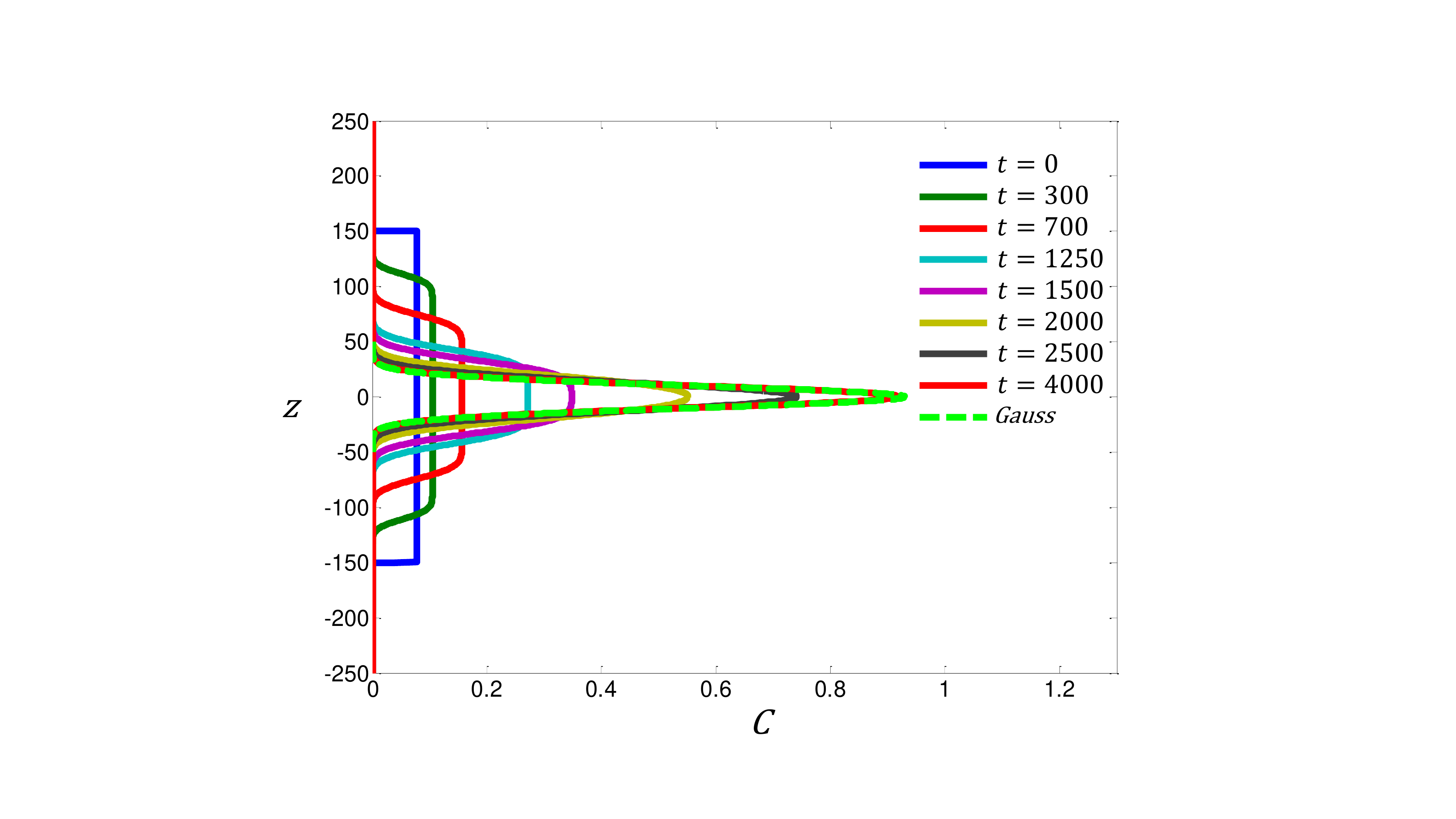}}
\caption{Evolution of the iron density profile in the simulation with $\Sigma_0 = 0.5 \Sigma_{\rm crit}$ described in the main text (solid lines), as well as the corresponding analytical prediction for the steady-state Gaussian solution  from equation (\ref{eq:laminarsteady}) for the same parameters. Note how the iron density profile tends to a Gaussian profile even though the initial condition (for this simulation only) are not Gaussian. }
\label{fig:S=0.5Scrit}
\end{figure}

In the case where $\Sigma_0 = 0.5 \Sigma_{\rm crit}$, we find that the system does indeed remain laminar. Furthermore, the iron density profile eventually converges to the steady-state laminar solution given in equation (\ref{eq:laminarsteady}) even when the initial conditions are not Gaussian (see Figure \ref{fig:S=0.5Scrit}).

\begin{figure}[h]
\centerline{\includegraphics[width=0.75\textwidth]{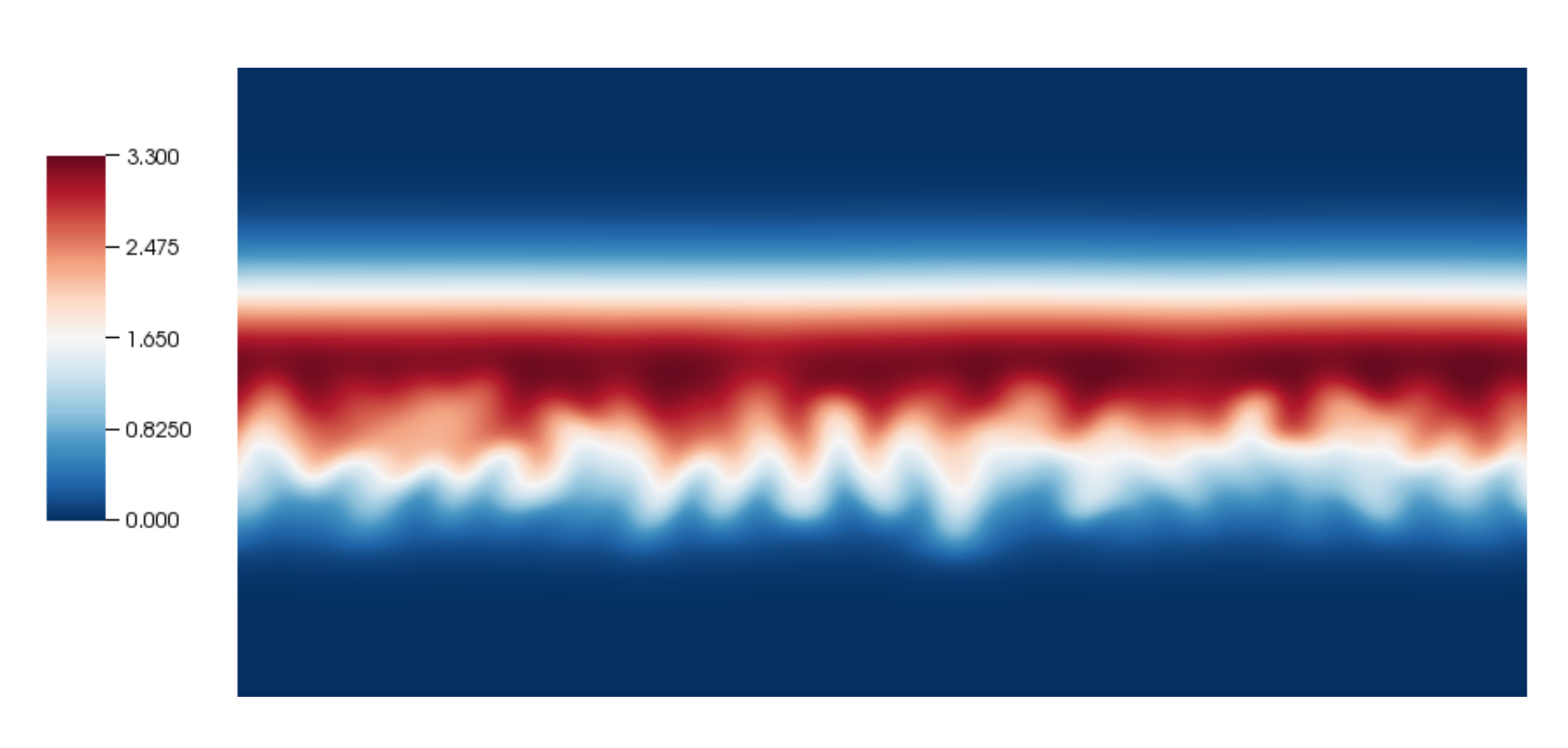}}
\caption{Snapshot of the iron density in the simulation with $\Sigma_0 = 2 \Sigma_{\rm crit}$ taken in the quasi-steady state reached after the onset of fingering convection. Fingering structures are clearly visible, and span most of the layer height. The layer asymmetry described in the main text, above and below $z=0$, is also clearly visible. }
\label{fig:S=2Scrit}
\end{figure}

In the case where $\Sigma_0 = 2 \Sigma_{\rm crit}$, by contrast, we find that fingering convection eventually takes place. Figure \ref{fig:S=2Scrit} shows a snapshot of the iron density once fingering has fully developed. The fingering structures, however, are about as tall as the iron layer height itself and are not very turbulent. This is not surprising since the system remains relatively close to the onset of instability at this value of $\Sigma_0$. 

\begin{figure}[h]
\centerline{\includegraphics[width=0.9\textwidth]{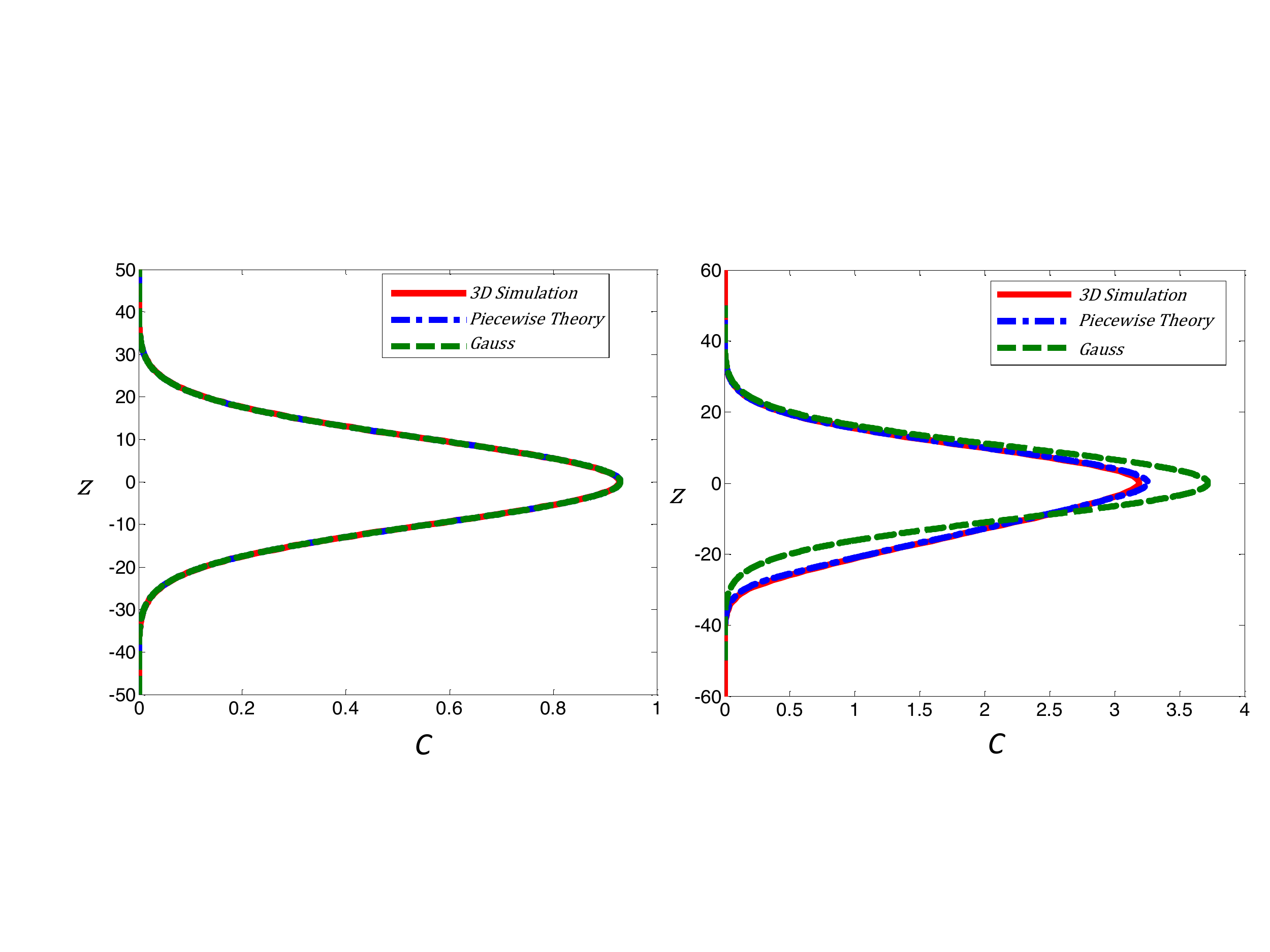}}
\caption{Steady state iron density profiles as a function of $z$. Left: case with $\Sigma_0 = 0.5 \Sigma_{\rm crit}$. Right: case with $\Sigma_0 = 2 \Sigma_{\rm crit}$. The solid red lines show the horizontally-averaged iron density profile from our 3D simulations, the dot-dashed blue line indicates the iron density profile from the piecewise analytical theory (see Section \ref{sec:piecewise}), and the dashed green line is the Gaussian laminar solution $C_\infty(z)$. While the latter is only a good fit to the true profile when the system remains laminar, the piecewise analytical theory fits extremely well in both cases.}
\label{fig:lowmass}
\end{figure}

In both cases, (i.e. for $\Sigma_0 = 0.5 \Sigma_{\rm crit}$ and for $\Sigma_0 = 2\Sigma_{\rm crit}$), the system eventually settles into a quasi-steady state in which the horizontally averaged iron density profile is independent of time. These are shown in Figure \ref{fig:lowmass}. In the laminar case, this steady state is given by equation (\ref{eq:laminarsteady}), and takes the form of a Gaussian of width $\Delta_\infty$. In the fingering case, the steady-state profile is no longer Gaussian, but is notably skewed downward. The upper part of the domain (above $z= 0$) is in balance between the upward down-gradient diffused flux, and the downward gravitational settling flux. The lower part of the domain is in balance between the upward levitated flux and the downward fingering flux. In Section \ref{sec:1D} we propose a simple model to explain the shape of this steady-state profile. This simple model predicts a steady-state profile (also shown in Figure \ref{fig:lowmass}) that clearly fits the results of our simulations remarkably well. 

\subsection{Intermediate mass regime}
\label{sec:intmass}

\begin{figure}
\centerline{\includegraphics[width=0.8\textwidth]{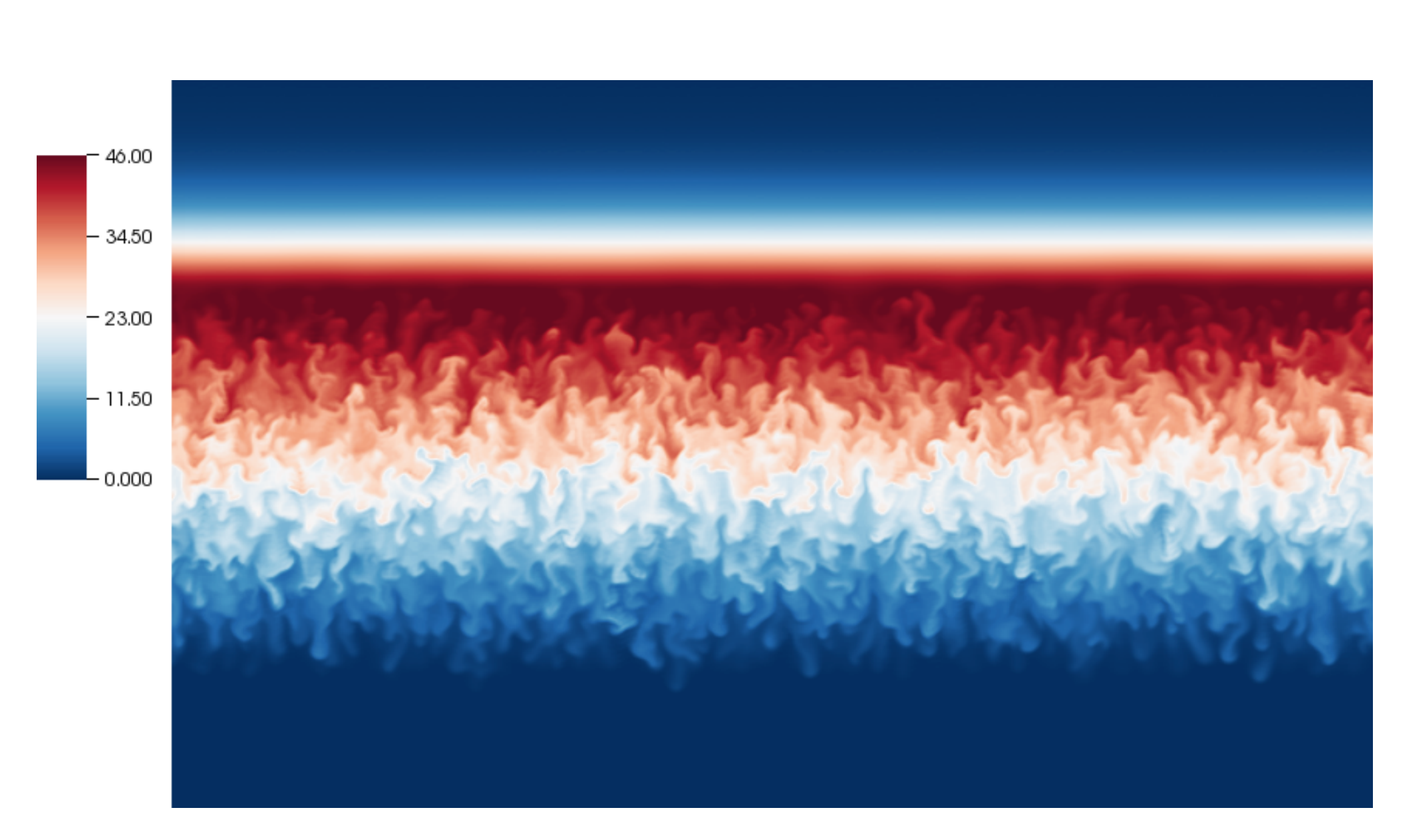}}
\caption{Snapshot of the iron density in the simulation with $\Sigma_0 = 100 \Sigma_{\rm crit}$ taken in the quasi-steady state reached after the onset of fingering convection. The fingering region is now more extended, asymmetric and turbulent than in the case where $\Sigma_0 = 2 \Sigma_{\rm crit}$.}
\label{fig:S=100Scrit}
\end{figure}

For larger values of $\Sigma_0$, we expect that the layer becomes unstable enough to have an extended turbulent fingering region. This is verified in Figure \ref{fig:S=100Scrit}, which shows a snapshot of a simulation with $\Sigma_0 = 100 \Sigma_{\rm crit}$, while all other parameters remain the same as in the previous section. The fingers are now clearly much more turbulent than before, and the layer extends over several finger scale heights (the latter being of the same order as the finger width). 

\begin{figure}
\centerline{\includegraphics[width=0.9\textwidth]{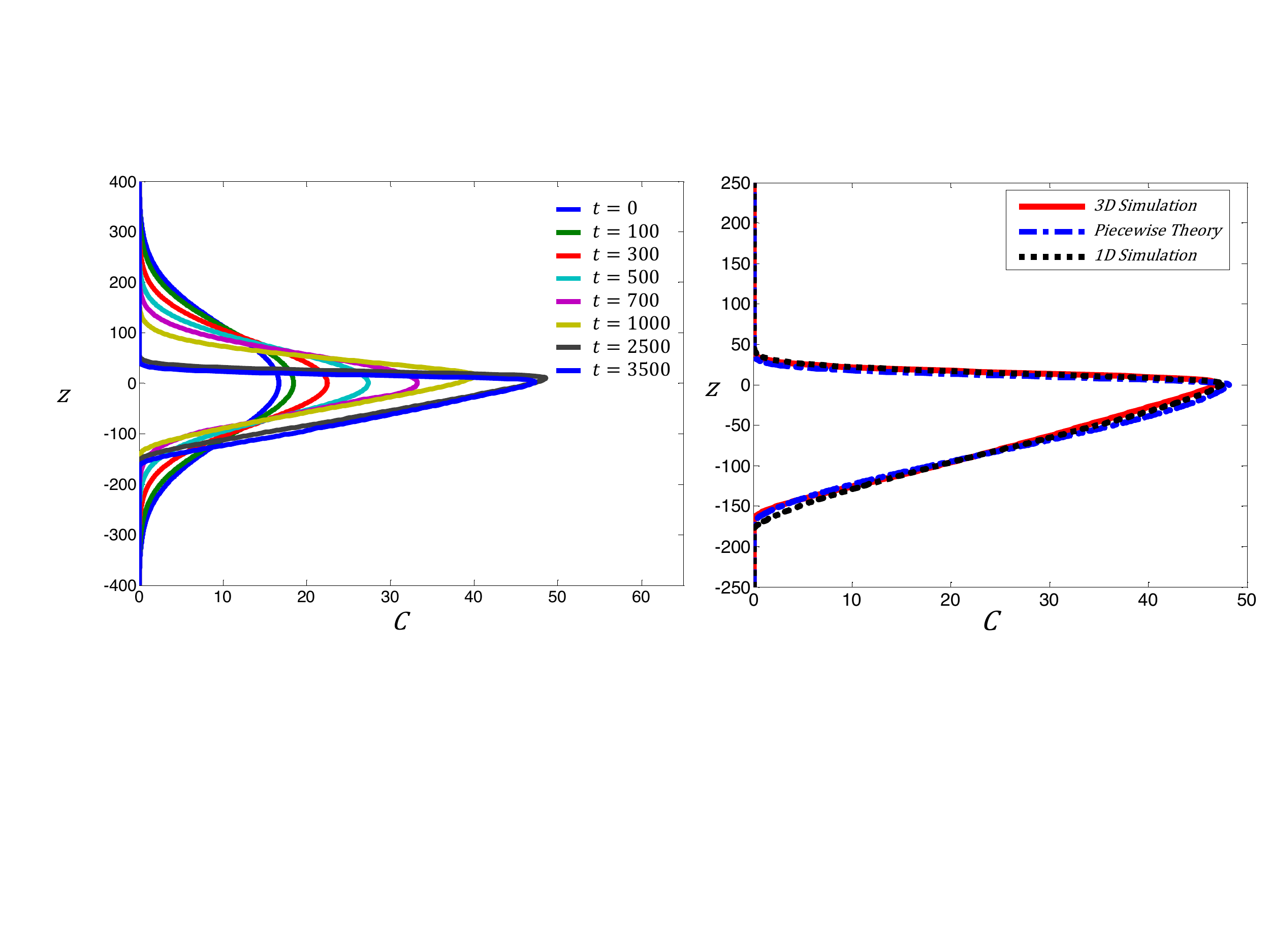}}
\caption{Left: Horizontally-averaged density profiles as a function of $z$ and time for systems with $\Sigma_0 = 100 \Sigma_{\rm crit}$. The concentration profile is initially stable, but becomes double diffusively unstable after contraction, and is no longer of Gaussian form at steady state. Right: Steady-state eventually achieved and comparison with our 1D models (see Section \ref{sec:1D}). The result of our piecewise semi-analytical model is shown in a blue dot-dashed line, while the steady-state profile achieved by evolving the 1D equation (\ref{turb_ad_diff}) together with the \citet{Brownal2013} parametrization of mixing by fingering convection in shown in the black dotted line.  }
\label{fig:highmass}
\end{figure}

Figure \ref{fig:highmass} shows the temporal evolution of the horizontally-averaged iron density profile, as well as its ultimate steady state. 
We find that the layer is now significantly asymmetric, and extends down to about $z=-150$. This is much larger than the laminar equilibrium thickness ($\Delta_\infty = 10$), but is comparable with the thickness at which the laminar Gaussian profile is expected to become turbulent (for these parameters, $\Delta_i = (\Sigma_0 / \Sigma_{\rm crit})  \Delta_\infty$ =100). 

\subsection{Fully-convective regime}
\label{sec:convreg}

\begin{figure}
\centerline{\includegraphics[width=0.3\textwidth]{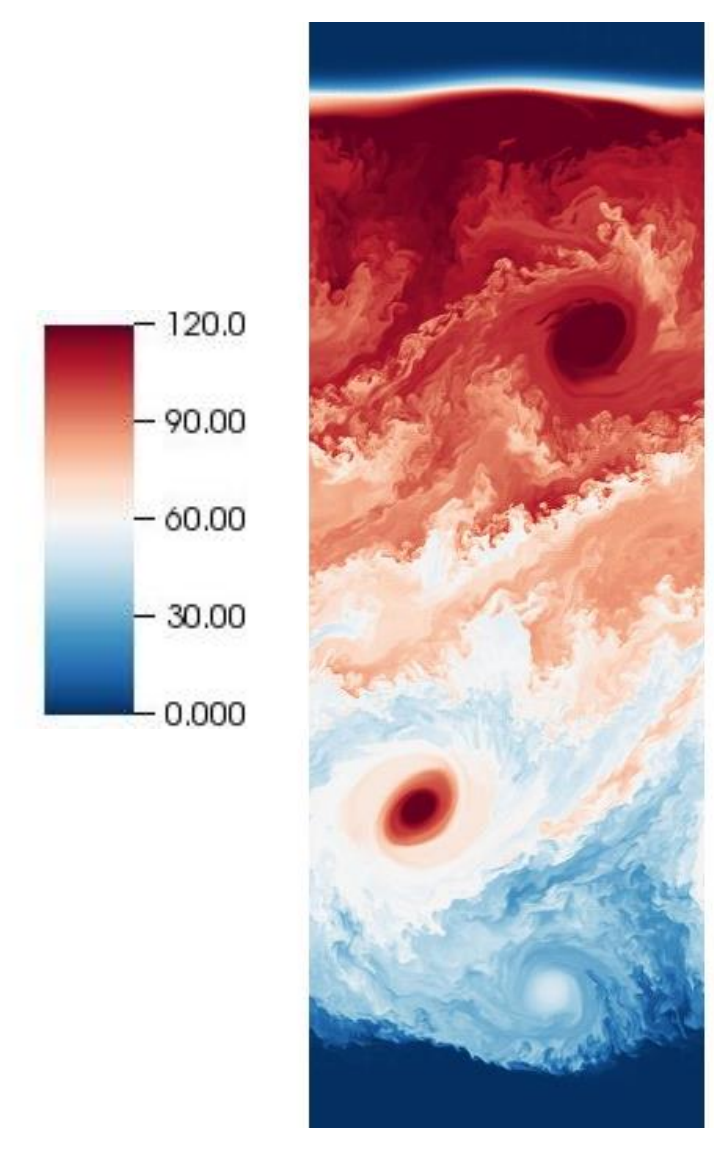}}
\caption{Snapshot of the iron density profile in the fully convective layer case, around $t = 4500$. The presence of strong iron concentrations in the convective eddies is partly an artefact of the aspect ratio of the computational domain used, which is quite thin. }
\label{fig:convsnaps}
\end{figure}

For even larger values of $\Sigma_0$, finally, we find that the compositional gradient that develops in the iron layer is so unstable that it becomes fully convective. By contrast with standard convective regions in stars, this kind of convection is (at least in the simulation) purely driven by the unstable iron density gradient, instead of being thermally-driven.
Figure \ref{fig:convsnaps} shows snapshots of the iron density for a simulation with $\Sigma_0 = 1000 \Sigma_{\rm crit}$, and for ${\rm Pr} = \tau =  0.1$ and $s=0.001$ as in previous cases. 
Note that with such a large value of $\Sigma_0$, $\Delta_i$ (and therefore $\Delta_{\rm init}$) is also quite large -- in this case, $\Delta_i = 100 \sqrt{10}$. For this reason, we had to use a very tall computational domain, with $z$ in the interval $[-1000,1000]$, to capture the entire initial Gaussian profile. The snapshots presented, however, zoom in on the region immediately around the convective layer\footnote{We actually ran two simulations: one in a thin domain (as in all the cases presented so far) of thickness $L_y = 15d$, and one in a thicker domain, of thickness $L_y = 64d$, to check whether the thin domain assumption may unnecessarily constrain the dynamics of the fully-convective layer. While we were not able to run the thick-domain model for as long as we have been able to run the thin-domain case, both models appear so far to be qualitatively, and statistically quantitatively similar in terms of compositional transport. We therefore focus our discussion to the thin-domain case for which we have a much longer dataset.}.

\begin{figure}
\begin{center}
\includegraphics[width=0.8\textwidth]{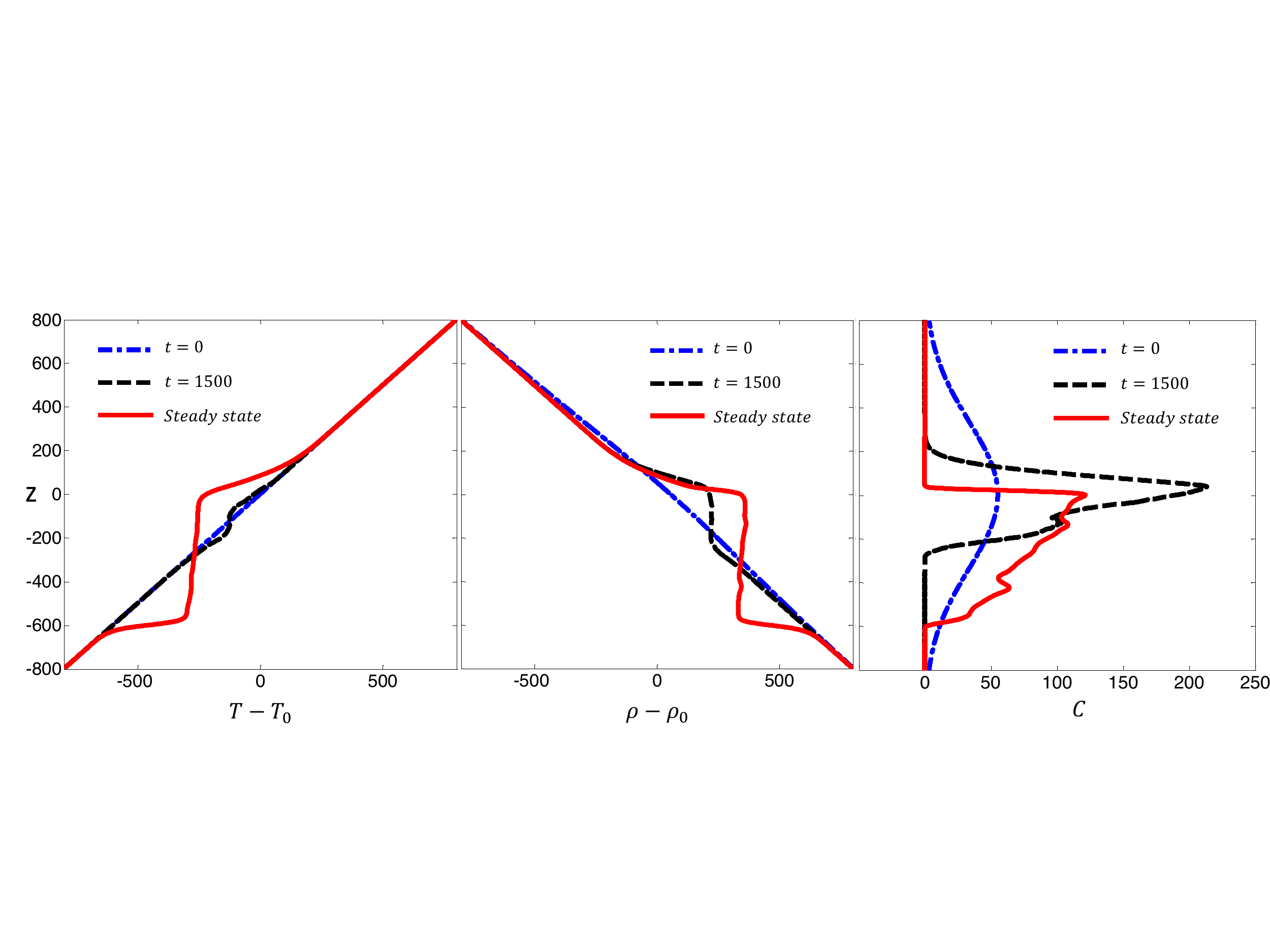}
\caption{Vertical profiles of the temperature and density deviations from their constant background values, as well as that of the iron density, for a simulation with $\Sigma_0 = 1000 \Sigma_{\rm crit}$, and for ${\rm Pr} = \tau =  0.1$ and $s=0.001$ as in the previous cases. Shown are profiles at times $t=0$ and $t = 1500$, as well as a short time averaged profile once the system has reached a steady state. The density increases very mildly with height within the layer as a result of the strong compositional anomaly, thereby maintaining a convective state. At steady state, the temperature profile is nearly uniform within the same region, but the iron density profile is not, decreasing smoothly from a peak around $z=0$ to zero near the bottom of the convective layer.   }
\label{fig:convprofs}
\end{center}
\end{figure}

It is quite clear from Figure \ref{fig:convsnaps} that the bulk of the iron layer now hosts large-scale eddies instead of individual fingers, in a manner that is strongly reminiscent of standard overturning convection. To prove that the layer is indeed fully convective, we show in Figure \ref{fig:convprofs} the vertical profiles of the temperature perturbation away from the mean temperature $T_0$, of the iron density, and of the density perturbation away from background density $\rho_0$, once the simulation has reached a stationary state. Both density and temperature profiles are clearly well-mixed in the layer, which roughly spans the interval $z \in[ -600, 0]$ once the system has reached a stationary state. The total density increases very mildly with height, as expected from a fully convective region. It is interesting to note that the concentration profile is not very well mixed by convection. This is not entirely surprising, in fact, since a fairly unstable compositional gradient is necessary to maintain an unstable density gradient against the stabilizing temperature stratification in this fairly peculiar kind of convection zone. 

In reality, however, this kind of compositionally-driven convection zone cannot exist at stellar parameter regimes in more realistic models (see below). Instead, as discussed by \citet{richard01}, convection zones are created by the increase in the local opacity as iron accumulates in the layer, an effect that is not modeled here.

\section{Simplified models for the long-term evolution of the system}
\label{sec:1D}

\subsection{A simple 1D model} 

The numerical simulations presented in Section 6 provide us with exact solutions to the set of idealized equations (\ref{eq:nondim}), and have revealed at least three distinct parameter regimes: the laminar regime (small surface density of iron $\Sigma_0$), the fingering regime (intermediate $\Sigma_0$), and the convective regime (large $\Sigma_0$). In each case, we were able to follow the evolution of the iron density profile from initial conditions to steady state, estimate the final layer width, turbulence level, and maximum iron density achieved. However, such simulations are not only computationally intensive, but can only be run for short periods of time compared with the stellar evolution timescale, in very small domains compared with the whole star, and at parameter values that are still very far from stellar. Furthermore, the governing equations are necessarily idealized, as discussed in the introduction, and neglect a number of phenomena such as the concurrent settling of helium, the feedback on the opacity caused by the accumulation of iron, and so forth. As a result, they should only be viewed as illustrative of the types of expected dynamics near the iron layer. 

However, as we now demonstrate, it is possible to create a simple 1D semi-analytical model of the evolution of the iron layer that both describes the results of the numerical simulations very accurately, can easily be extended to stellar parameter values, can straightforwardly be included in a stellar evolution code, and can thus be generalized to include the aforementioned missing physics such as helium settling or a temperature and composition-dependent opacity. To do so, we make use of the recent work of \citet{Brownal2013}, who proposed a model for turbulent transport by fingering convection in the stellar parameter regime.   


The evolution of layers that become fingering-unstable is controlled by the sum of the down-gradient diffused element flux, of the levitated flux (from below), of the settling flux (from above) and finally of the fingering or convective flux (in the turbulent layer below $z=0$). The first three are purely laminar processes, and already taken into account in equation (\ref{eq:ad_difflaminar}) for instance. To model the effect of the last one, we now consider a slightly modified 1D advection-diffusion equation that includes the turbulent flux $F_C$:  
\begin{equation}
\frac{\partial C}{\partial t}  -  \frac{\partial}{\partial z} (szC) + \frac{\partial F_C}{\partial z} = \tau \frac{\partial^2 C}{\partial z^2} \mbox{  .} 
\label{turb_ad_diff}
\end{equation}
Limiting our analysis to the case of fingering convection only, we use the results of \citet{Brownal2013} who showed that $F_C$ can be expressed in terms of a Nusselt number, ${\rm Nu}_C$, defined such as:\\
\begin{equation}
F_C = - \tau({\rm Nu}_C - 1) \frac{\partial C}{\partial z} \mbox{  ,} 
\label{eq:fluxeq}
\end{equation}
where ${\rm Nu}_C$ depends on the local values of ${\rm Pr}$, $\tau$ and of
\begin{equation}
R(z,t) = \left( \frac{\partial C}{\partial z}  \right)^{-1} \mbox{   ,}
\label{eq:Rdef}
\end{equation}
the local density ratio at the position considered.
Combining (\ref{eq:fluxeq}) with (\ref{turb_ad_diff}), we then get
\begin{equation}
\frac{\partial C}{\partial t}  - s \frac{\partial}{\partial z} (z C) = \tau \frac{\partial}{\partial z} \left( {\rm Nu}_C (z,t)  \frac{\partial C}{\partial z}  \right).
\label{turb_ad_diff2}
\end{equation}

\citet{Brownal2013} showed that the compositional Nusselt number takes the form:\\
\begin{equation}
{\rm Nu}_C(z,t) = 1 + 49 \frac{\lambda^2}{\tau l^2 (\lambda + \tau l^2)}\mbox{   ,}
\label{eq:Nubrown}
\end{equation}
where $\lambda$ and $l$ are the growth rate and the wavenumber of the fastest growing mode of the fingering instability at the parameters ${\rm Pr}$, $\tau$ and $R(z,t)$. As such, they both depend on $z$ and $t$ as well, and are found from a standard linear stability analysis described in Appendix A of \citet{Brownal2013}. Note that if the layer is stable to fingering convection then we set ${\rm Nu}_C=1$, so $F_C = 0$ and equation (\ref{turb_ad_diff}) recovers the original advection-diffusion equation (\ref{eq:ad_difflaminar}). 

\subsection{Comparison with 3D simulations}
\label{sec:1dvs3d}

To check the validity of this simple 1D model, we compare its evolution to that of the full 3D simulations presented in Section \ref{sec:num}. 
We compare for instance the temporal evolution of the iron density profile for the case with $\Sigma_0 = 100 \Sigma_{\rm crit}$,
using exactly the same initial conditions for the two runs. We have checked that the profiles match throughout the course of the system's evolution, and show in Figure \ref{fig:highmass}b a comparison between the ultimate steady states of the 1D and 3D models. The match between the two solutions is of course not perfect, but is nevertheless incredibly good 
given the highly-simplified nature of the 1D model. Also note that the 1D model contains no free parameters. 
This result ultimately confirms that the \citet{Brownal2013} model can be safely used in models of stellar evolution with atomic diffusion to represent the effects of fingering on compositional transport.

\subsection{Steady-state models for fingering systems.}
\label{sec:piecewise}

Having established that the 1D equation (\ref{turb_ad_diff2}) is a good approximation to the true 3D behavior of the system, we can now use it 
to find what the ultimate quasi-steady state of the iron layer may be while undergoing active fingering convection for a wide range of input parameters. To do so, we must solve:
\begin{equation}
-szC = \tau {\rm Nu}_C(z) \frac{\partial C}{\partial z}.
\label{ss_eqn}
\end{equation}
Although this equation looks simple, it is in fact highly nonlinear since ${\rm Nu}_C$ is a rational function of both $\lambda$ and $l$, each of which is a complicated 
function of $\frac{\partial C}{\partial z}$ via $R(z,t)$. 

There are two possible ways of solving (\ref{ss_eqn}) directly (i.e. without proceeding through an expensive time-dependent evolution first, as we have in Section \ref{sec:1dvs3d}): 
one can either do so numerically using a two-point boundary value solver for nonlinear ordinary differential equations, 
or, one can try to approximate (\ref{ss_eqn}) so that it lends itself to a simpler analytical treatment. In what follows, we pursue the second option, in order to try to develop more intuition about the general behavior of the iron layer at different parameter regimes. 
\begin{figure}
\begin{center}
\includegraphics[width=4in]{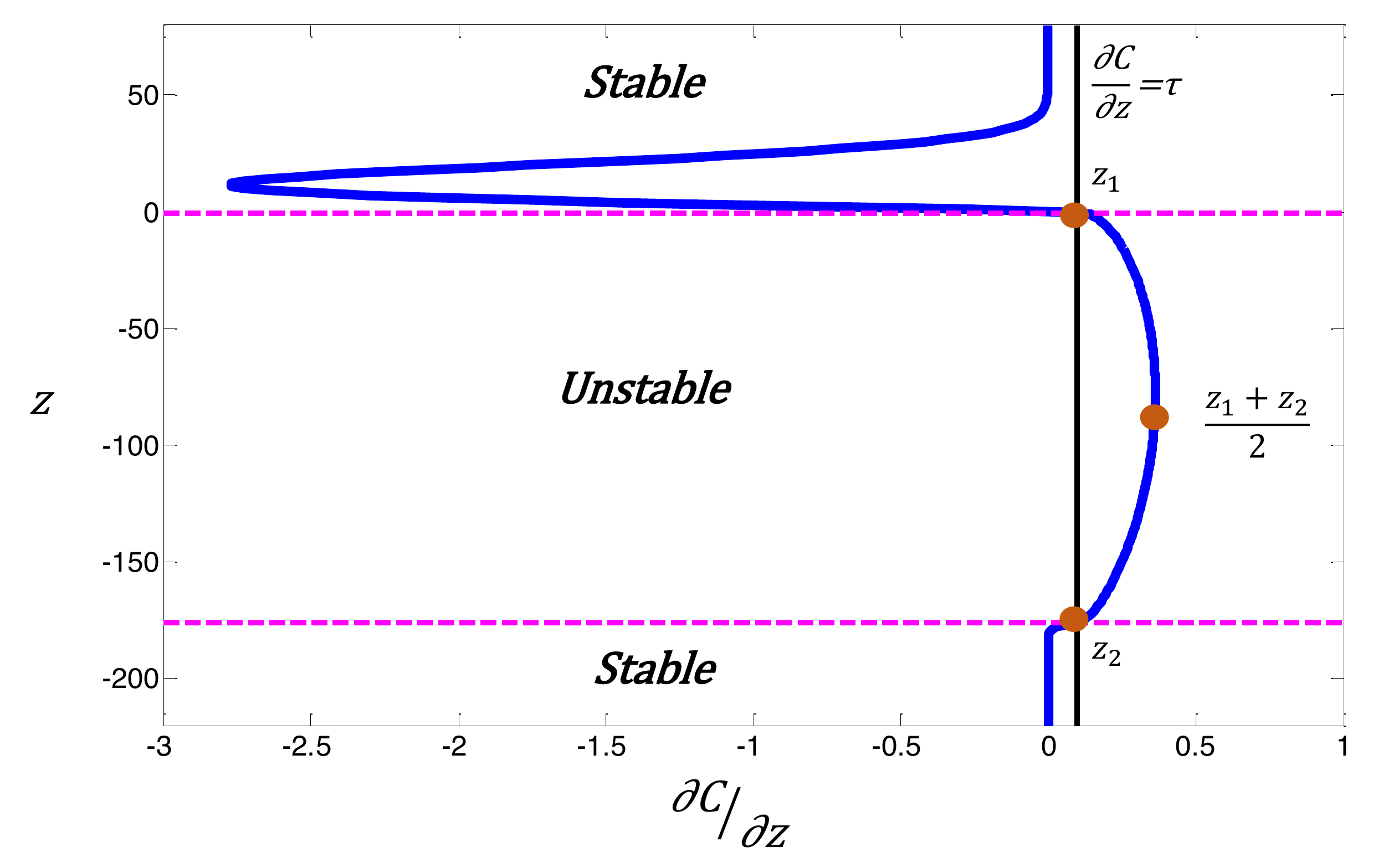}
\caption{The blue solid line shows the vertical derivative of the steady-steate iron density profile $\frac{\partial C}{\partial z}$, for the parameters ${\rm Pr } = \tau = 0.1$, $s = 0.001$ and $\Sigma_0 = 100 \Sigma_{\rm crit}$. The black vertical line lies at the constant value $\tau$. This profile illustrates various regions of interest: regions where $\frac{\partial C}{\partial z} < \tau$ (for $z>z_1$ and $z<z_2$) are stable to fingering convection, while the region with $\frac{\partial C}{\partial z} > \tau$ (for $z \in [z_2,z_1]$) are fingering unstable. Note that the $\frac{\partial C}{\partial z}$ profile is close to parabolic in that region. }
\label{th_sch}
\end{center}
\end{figure} 


To do so, we first inspect the numerical solution obtained in Section \ref{sec:intmass} for the case where $\Sigma_0 = 100 \Sigma_{\rm crit}$ in more detail. Figure \ref{th_sch} shows the horizontally-averaged profile of $\frac{\partial C}{\partial z}$ as a function of $z$ for the ultimate steady-state profile achieved. The profile clearly has three distinct parts: above $z_1$ and below $z_2$, $\partial C/\partial z < \tau$, so the system is stable to fingering convection. This means that ${\rm Nu}_C = 1$, and that the overall flux balance is between diffusion and settling (for $z> z_1$) or diffusion and levitation (for $z<z_2$). Between $z_1$ and $z_2$, however, $\tau < \partial C/\partial z < 1 $, so the layer is fingering-unstable. There, ${\rm Nu}_C > 1$ and is given by equation (\ref{eq:Nubrown}). Interestingly, we can see that the profile for $\partial C/\partial z$ is close to being parabolic between $z_1$ and $z_2$. These considerations suggest the following construction for an approximate steady-state solution.

Above $z_1$, and below $z_2$, the steady-state solution satisfies (\ref{eq:ad_difflaminar}), which was already solved in Section \ref{sec:laminar} and found to be a Gaussian profile. 
The main difference is that the solutions in the two regions may now have different normalization constants, so that 
\begin{eqnarray}
&& C(z \geq z_1) = \frac{\Sigma_1}{\sqrt{2 \pi} \Delta_{\infty}} e^{-\frac{z^2}{2 \Delta_{\infty}^2}} \mbox{   ,} \nonumber \\
&& C(z \leq z_2) = \frac{\Sigma_2}{\sqrt{2 \pi} \Delta_{\infty}} e^{-\frac{z^2}{2\Delta_{\infty}^2}} \mbox{   ,} 
\end{eqnarray}
where $\Delta_\infty$ was defined in (\ref{eq:laminarsteady}), and $\Sigma_1$ and $\Sigma_2$ are two constants to be determined from global mass conservation (see below). 

Since $z_1$ and $z_2$ are by construction points of marginal stability for fingering convection, we both have  $\partial C/\partial z = \tau$ and ${\rm Nu}_C = 1$ at these points. This then implies 
\begin{equation}
-sz_1C(z_1) = \tau \left. \frac{\partial C}{\partial z}\right|_{z_1} = \tau^2 , \mbox{  and  }  -sz_2C(z_2) = \tau  \left. \frac{\partial C}{\partial z}\right|_{z_2}
 = \tau^2 \mbox{   .} 
\end{equation}
Hence we find that $C(z_1) = - \tau^2 / s z_1$ and $C(z_2) = -\tau^2 /s  z_2$, which then yields $\Sigma_1$ and $\Sigma_2$ provided $z_1$ and $z_2$ are known: 
\begin{eqnarray}
&& \Sigma_1 = -\sqrt{2 \pi }  \frac{\Delta_{\infty} \tau^2}{sz_1} e^{\frac{{z_1}^2}{2 \Delta^2_{\infty} }}  \mbox{   ,}  \nonumber \\
&& \Sigma_2 = -\sqrt{2 \pi }  \frac{\Delta_{\infty} \tau^2}{sz_2}  e^{\frac{{z_2}^2}{2 \Delta^2_{\infty} }} \mbox{   .} 
\end{eqnarray}
So far, all of these solutions and algebraic equations relating the various variables are exact. However, they do not yet form a closed system, since we do not know a priori what $z_1$ and $z_2$ are. The dynamics of the intermediate fingering region, for $z \in [z_1,z_2]$, is of course crucial in closing the problem, but is more difficult to solve exactly.

For this reason, we simply approximate $\partial C/\partial z$ in that region using a parabola: $\partial C/\partial z = \tau + k (z_1 - z)(z-z_2)$. This introduces a new constant, $k$, so we now have three unknowns to solve for ($k$, $z_1$ and $z_2$). The latter are given by the solution of the following three equations: 
\begin{equation}
\Sigma_0 + \frac{\tau \Delta_\infty^3}{z_1} \sqrt{\frac{\pi }{2}} e^{\frac{{z_1}^2}{2 \Delta^2_{\infty}}}  \left[1 - {\rm erf}\left(\frac{z_1}{\sqrt{2} \Delta_{\infty}}\right)\right] + \frac{\tau \Delta_\infty^3}{z_2} \sqrt{\frac{\pi }{2}} e^{\frac{{z_2}^2}{2 \Delta^2_{\infty}}} \left[1 + {\rm erf}\left(\frac{z_2}{\sqrt{2} \Delta_{\infty}}\right)\right] = 0  \mbox{   ,}
\label{ss_crit1}
\end{equation}
\begin{equation}
 \tau =  \frac{z_1 z_2}{\Delta_\infty^2}  \left[\tau + \frac{k}{6} (z_1 - z_2)^2\right]   \mbox{   ,}
\label{ss_crit2}
\end{equation}
\begin{equation}
\frac{(\tau z_1 + z_2)^2}{4 z_1 z_2} - {\rm Nu}_C (C_z(\hat{z})) \left[\tau + \frac{k (z_1-z_2)^2}{4}\right] = 0  \mbox{   ,}
\label{ss_crit3}
\end{equation}
where $\hat{z} = (z_1+z_2)/2$. Equation (\ref{ss_crit1}) is obtained from mass conservation by integrating the approximate piecewise iron density profile from $-\infty$ to $z_2$, then from $z_2$ to $z_1$, and finally from $z_1$ to $+\infty$. Equation (\ref{ss_crit2}) is found by matching the diffusive to the fingering solutions at $z_1$ and $z_2$. Finally, Equation (\ref{ss_crit3}) is obtained from equation (\ref{ss_eqn}) at  $\hat z$, and provides a constraint on $k$. 
 
\noindent These three equations can be solved using Newton's method for a given set of input parameters: $\tau$, $\rm{Pr}$, $s$, and $\Sigma_0$. Once these are known, the piecewise steady state iron density profile $C(z)$ is obtained from:
\begin{eqnarray}
&& C(z > z_1) = -\frac{\Delta_{\infty}^2 \tau}{z_1} e^{-\frac{z^2 - {z_1}^2}{2 \Delta_{\infty}^2 }},
\label{pwC_1} \nonumber \\
&& C(z_2 \leq z \leq z_1) = -\frac{\Delta_{\infty}^2  \tau}{z_2}  + \tau(z-z_2) + \frac{k}{2}(z_1 - z_2)(z-z_2)^2 - \frac{k}{3}(z-z_2)^3 ,
\label{pwC_2} \nonumber \\
&& C(z < z_2) = -\frac{\Delta_{\infty}^2  \tau}{z_2} e^{-\frac{z^2 - {z_2}^2}{2 \Delta_{\infty}^2 }}.
\label{pwC_3}
\end{eqnarray}

\begin{figure}
	\centerline{\includegraphics[width=0.5\textwidth]{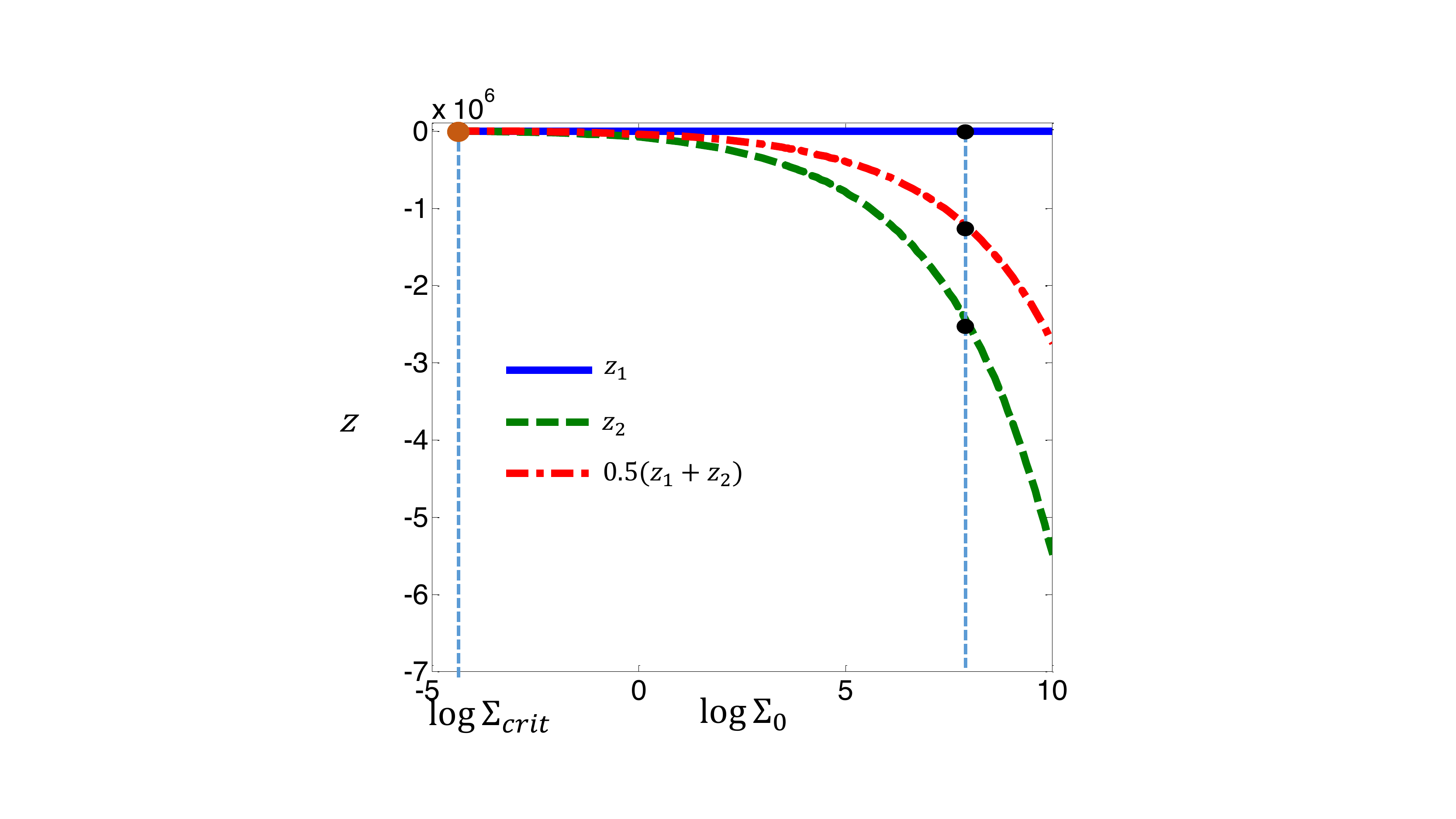}}
	\caption{Variation of the upper and lower extent of the iron layer, $z_1$ and $z_2$, as functions $\Sigma_0$, for a system with stellar input parameters (see Table 2), as calculated using our piecewise algebraic method (see Section 7.3). The vertical dashed line at $\Sigma_0 \sim 10^8$ corresponds to the expected value of $\Sigma_0$ in the stellar model described in Tables 1 and 2.}
\label{turb_massincrease}
\end{figure}

A comparison of the exact profiles from 3D simulations, of the 1D profile from the integration of the 1D model, and of the piecewise algebraic model, is presented in Figure \ref{fig:highmass}b for $ {\rm Pr} = \tau = 0.1$, $s=0.001$, and $\Sigma_0 = 100\Sigma_{\rm crit}$. Another comparison for $\Sigma_0 = 2\Sigma_{\rm crit}$ is shown in Figure \ref{fig:lowmass}. In both cases, the fit is excellent. 

The advantage of the piecewise algebraic method is that solutions can be obtained very rapidly for any value of the input parameters, including realistic stellar values. 
Applying it to the parameters given in Tables 1 and 2, we then get the results presented in Figure \ref{turb_massincrease}.
This suggests that  the typical steady-state iron layer width, for reasonable input values for the total mass of iron (corresponding to $\Sigma_0 \sim 10^8$ in non-dimensional units, see Table 2) is about $3\times 10^6 d$, which in dimensional terms corresponds to about $10^{12}$cm, or in other words, a width larger than the radius of the star itself! 


These results, taken at face value, imply that mixing by fingering convection in the absence of all other processes (see below) is {\it so} efficient in this parameter regime that iron should eventually be redistributed throughout the entire star, and therefore cannot accumulate in a thin layer or lead to the formation of an iron convection zone, at least in a steady state situation. Of course, reaching a steady state could take a very long time, and temporarily thinner layers can in principle be found. One could study this using the time-dependent 1D approach presented in Section \ref{sec:1dvs3d}. However, as discussed by \citet{theado09} and \citet{richard01}, other important processes that were neglected in our 3D calculation {\it must} be taken into account to properly model the evolution of the iron layer anyway. It is now time to go back to full stellar evolution models that take into account all relevant physics, albeit in one-dimension only, to complete our study.

\section{Application to stellar evolution models}

Having established that the prescription for turbulent mixing by fingering convection proposed by \citet{Brownal2013} correctly accounts for the 3D dynamical evolution of an ``idealized" iron layer, we now apply it to full stellar evolution calculations. We revisit the model 1.7 M$_\odot$ star discussed in Section 2, and add the computation of fingering convection to that of atomic diffusion in order to follow the evolution of the iron abundance profile with time.

At each time step and each radial position in the star, the stellar evolution calculation yields the actual realistic values of non-dimensional parameters such as Pr, $\tau$ and the local density ratio $R(r,t)$ (see equation \ref{eq:densityratio}), which are then used to determine whether the local inverse $\mu$-gradient is sufficiently strong to trigger fingering convection, and if so, to calculate the expected fingering flux according to equation (\ref{eq:fluxeq}). The contribution of this turbulent flux is then added to the equation for the evolution of the local iron abundance, and the latter is updated accordingly at the next time step.



By contrast with the necessarily simplified model used in our 3D simulation, the stellar evolution computations include all relevant physics, and in particular that of helium settling, which can create a stabilizing $\mu$-gradient in competition with the destabilizing one induced by the iron accumulation \citep{theado09}. It also includes the dependence of the local opacity on the iron abundance, which could play an important role -- according to \citet{richard01} -- in triggering standard overturning convection. In order to gain a better understanding of the results of the simulations, we first computed toy models in which helium diffusion is artificially suppressed, and then computed more realistic models that include it.


\subsection{Toy stellar model without helium settling}

\begin{figure}
\begin{center}
\includegraphics[width=4.5in]{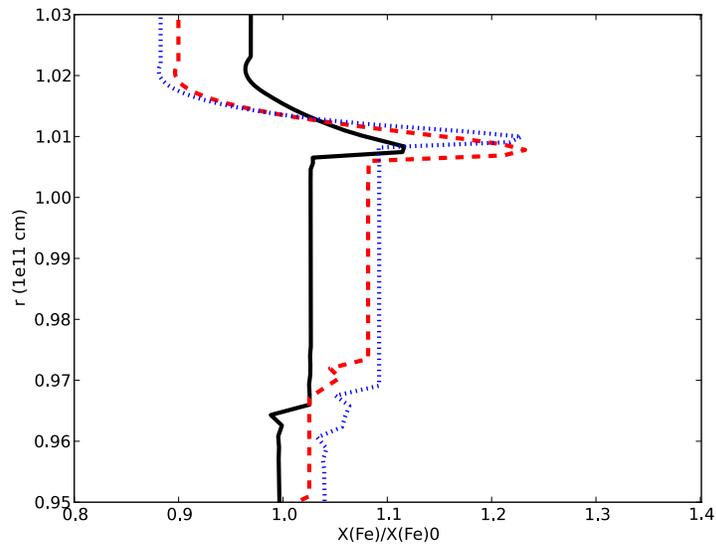}
\caption{Iron abundance profiles in a 1.7 M$_{\odot}$ star at 23Myrs (black solid line), 28Myrs (red dashed line) and 35 Myrs (blue dotted line) in models with fingering convection, but without helium settling. This figure can be directly compared with Figure \ref{Fe-noth}. Note how the overabundance of iron this time remains quite small.  }
\label{Fe-th}
\end{center}
\end{figure}




The iron abundance profiles in the toy model without helium settling are presented in Figure \ref{Fe-th} for the same ages (23Myrs, 28 Myrs, and 35 Myrs) as the model without fingering described in Section 2. The profile contains a peak around $r_0 = 1.01 \times 10^{11}$cm. Below the peak lies a very efficiently mixed fingering region where the iron abundance is homogeneous. The differences between these profiles and the ones presented in Figures \ref{fig:highmass} and \ref{fig:convsnaps} is striking, but can easily be explained by results of Section \ref{sec:piecewise} and the fact that Pr and $\tau$ are five or six orders of magnitude smaller in the real stellar case than the parameters used in the 3D simulations. At these realistic stellar parameters, the \citet{Brownal2013} model predicts vastly increased turbulent fluxes compared with the ones found in the 3D simulations at higher Pr and $\tau$, and the larger turbulent iron diffusivity leads to a much wider, and more efficiently mixed iron layer. Presumably, this layer will continue to thicken slowly with time as mixing by fingering convection proceeds. 

Comparing Figures \ref{Fe-noth} and \ref{Fe-th}, we see that mixing by fingering convection strongly reduces the iron over-abundance in the layer. The case with fingering convection (but without helium settling) has an overabundance of about 20\% at most, while that without fingering altogether has eight times more iron in the layer than in the surrounding regions. As a further result, we also find that the iron density in fingering layers never becomes large enough for opacity effects to be important and trigger the onset of overturning convection.


\subsection{Realistic stellar models with helium settling}

In this section, we present a more realistic evolution model of the same 1.7 M$_\odot$ star, including this time the diffusion of all the elements, and notably that of helium as well as iron, as described in Section 2. For reasons that will be clarified shortly, we evolve these more realistic models for much longer than the one presented in Section 2 and above, up to 500 Myrs. The iron accumulation initially occurs at roughly the same radius in this new model as it did in the toy model, but changes radius as a consequence of the global evolution of the star on these longer timescales. For this reason, we now show results in Lagrangian mass coordinates rather than radial coordinates. The iron layer remains roughly at the same mass coordinate with time, thus facilitating the analysis of its evolution. The characteristics of the stellar background in the iron layer region are similar, albeit not identical of course, to those of Table 2.


\begin{figure}
\begin{center}
\includegraphics[width=4.5in]{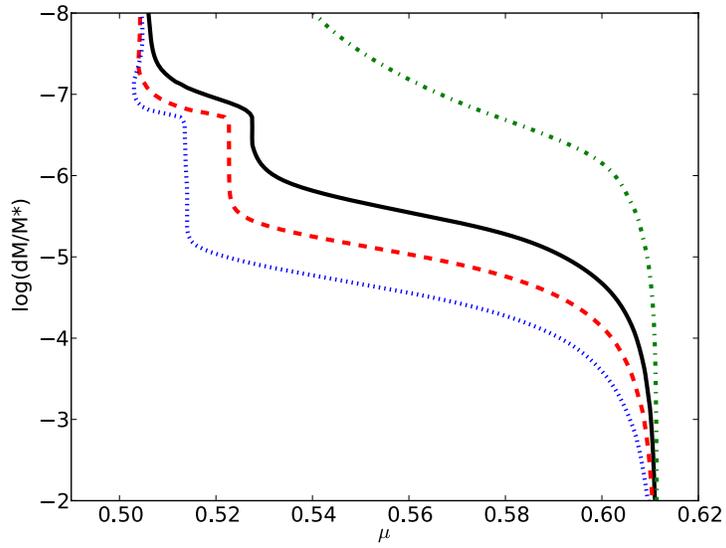}
\caption{$\mu$-profiles computed using the full realistic stellar evolution model for a 1.7 M$_\odot$ star at 35Myrs (green dot-dashed line), 123Myrs (black solid line), 243Myrs (red dashed line) and 490 Myrs (blue dotted line), as a function of the fractional stellar mass below the photosphere. Note that the $\mu$-profile contains a barely visible inversion at the latest time only, where $\log dM/M_* \simeq -5.5$, which is unstable to fingering convection. At early times the $\mu$-gradient is strongly stabilizing against fingering convection. }
\label{mu-real}
\end{center}
\end{figure}

\begin{figure}
\begin{center}
\includegraphics[width=4.5in]{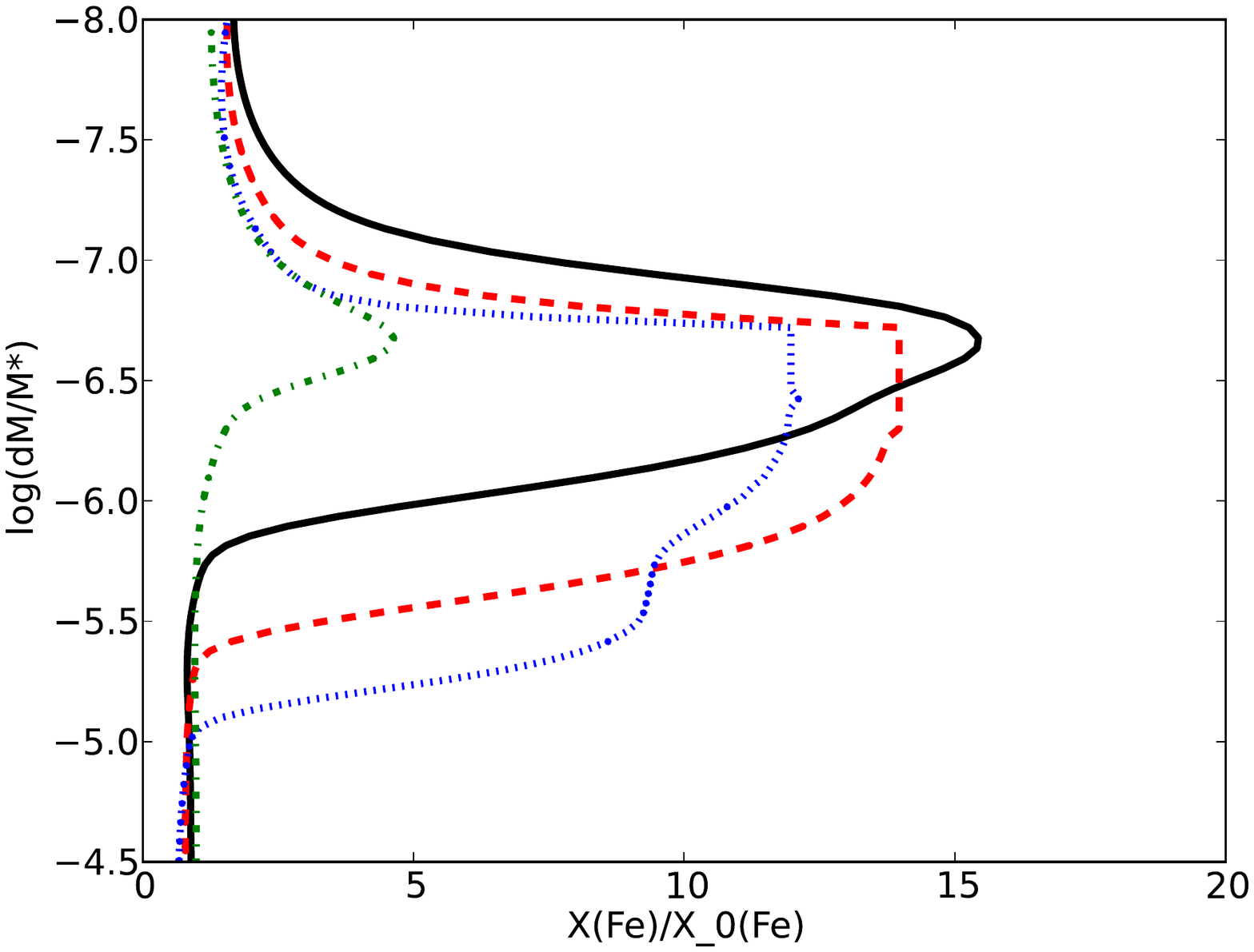}
\caption{Iron profiles computed using the full realistic stellar evolution model for a 1.7 M$_\odot$ star at 35Myrs (green dot-dashed line), 123Myrs (black solid line), 243Myrs (red dashed line) and 490 Myrs (blue dotted line), as a function of the fractional stellar mass below the photosphere. The effect of fingering convection only appears at 490 Myrs. Before that age, it is prevented by the helium gradient. The different shapes at 123 and 243 Myrs are due to the occurence of opacity-induced dynamical convection.}
\label{Fe-th-real}
\end{center}
\end{figure}

Figure \ref{mu-real} shows the overall $\mu$-profile at various ages from 35Myrs to 500Myrs in the vicinity of the iron layer. We see that a strongly stabilizing $\mu$-gradient rapidly develops as a result of helium settling. This gradient is in fact sufficiently strong to completely stabilize the system against fingering convection, allowing a significant amount of iron to accumulate in the layer as shown in Figure \ref{Fe-th-real}. At 35Myrs (green curve), the profile is very close to being Gaussian and there is not mixing present. It is only much later (after 300 Myrs) that the iron over-abundance becomes sufficiently large to overwhelm the helium-induced $\mu-$gradient, to form a mild {\it inverse} $\mu$-gradient and drive fingering convection.

We clearly see that the iron abundance in the peak may become much larger than for the toy model without helium settling, showing that the latter {\it must} be taken into account to properly model the formation and evolution of iron layers in stars. In fact, the iron abundance in this case becomes sufficiently large as to increase the local opacity significantly, and thus induce the formation of a thermally-driven convection zone (by contrast with the compositionally-driven convection zone discussed in Sections 6.4 and 7.4). This first happens around 100Myrs. The black and red curves, which are displayed for two ages, 128 and 243 Myrs, both show the presence of a mixed region that corresponds to these convective zones. 

We find that the iron layer in this simulation oscillates between convective and non-convective regimes during stellar evolution, as already shown by \citet{theado09} (see their figure 8). When fingering convection finally begins (around 300Myrs), it first sets in directly below the convective region, and drains the excess iron from the radiative--convective interface downwards. This modifies the profile's shapes further, extending the homogeneously mixed region significantly, and gradually decreasing the maximum value of the iron peak with time (see profile around 490Myrs, dotted blue curve). In these models again, we do not find that the iron profile reaches a steady state, but instead, continues to change with time throughout the Main Sequence evolution of the star.

\section{Summary and conclusion }

In this work, we presented a comprehensive study of the dynamics of iron layers that commonly form in the interior of A-type stars, due to atomic diffusion.
The presence, nature and structure of these layers can have important effects on the properties of stellar pulsations \citep{richard01,theado09}, and must therefore be better understood in order to utilize the 
asteroseismic results of the {\it Kepler} mission to their full potential.  

As shown by \citet{theado09}, iron layers are intrinsically prone to fingering instabilities that develop whenever an unstable $\mu$-gradient appears. This naturally regulates the amount of iron 
allowed to accumulate, and is a process that must be modeled correctly to obtain quantitatively robust predictions for the layer structure.  
Using 3D idealized numerical simulations of such an iron layer, we first tested the recent prescription proposed by \citet{Brownal2013} for mixing by fingering convection. 
This prescription contains no free parameters, and we found it to be an excellent
model, able to reproduce the horizontally-averaged iron density profile within the layer to within a few percent. 

Using this model in conjunction with the Toulouse-Geneva Evolution code, we then compared different evolution scenarios for the iron layer in order to better understand what role each process plays. 
Taking into account the atomic diffusion of {\it all elements}, but ignoring fingering convection, leads to vastly over-estimated predictions for the iron over-abundance in the layer, 
and under-estimated predictions for its thickness. Conversely, taking into account the atomic diffusion of {\it iron only}, but with fingering convection, we find that the iron layer is strongly diluted downward by the fingering-induced mixing, becomes very extended but only supports a very weak iron over-abundance. Finally, taking all processes properly into account, we find that helium settling is a crucial component of the dynamics of the system, since it strongly stabilizes the layer against fingering instabilities. Iron is able to accumulate significantly for several hundred thousand years before fingering convection finally sets in. Even when it does, it is not as efficient as in the case without helium settling, and significant overabundances can be preserved. As a result, we also find that the increase in the local opacity caused by the increased iron content can trigger thermal convection, as discussed by \citet{richard01}.

These results show that studying iron-accumulation layers requires modeling a combination of a number of processes accurately. Thankfully, we now have all the tools to do so in a well-tested, parameter-free way, and will in the future be applying our model to specific stars, in view of comparing their predicted pulsation properties with {\it Kepler} observations. 

\acknowledgements

This work originated from Varvara Zemskova's summer project at the Woods Hole GFD Summer Program in 2013. We thank the NSF and the ONR for supporting this excellent program.
We also thank Eckart Meiburg for many useful discussions during the early stages of Varvara Zemskova's project. We thank Sylvie Th\'eado who implemented the computations of radiative accelerations in the Toulouse-Geneva Evolution Code

\addtocontents{toc}{\protect\vspace*{\baselineskip}}

\bibliography{DDC_bib_TV4}



\appendix


\end{document}